\documentclass[aps, superscriptaddress, longbibliography, notitlepage, nofootinbib]{revtex4-1}

\usepackage{amsmath,amssymb, mathtools}
\usepackage{graphicx}
\usepackage[colorlinks=true,allcolors=blue]{hyperref}
\usepackage{enumitem}

\begin{document}

\title{Hyperbolic Mapping of Human Proximity Networks}

\author{Marco A. Rodr\'{i}guez-Flores}\email{mj.rodriguezflores@edu.cut.ac.cy}
\affiliation{Department of Electrical Engineering, Computer Engineering and Informatics, Cyprus University of Technology, 3036 Limassol, Cyprus}
\author{Fragkiskos Papadopoulos}\email{f.papadopoulos@cut.ac.cy}
\affiliation{Department of Electrical Engineering, Computer Engineering and Informatics, Cyprus University of Technology, 3036 Limassol, Cyprus}

\date{\today}

\begin{abstract}
Human proximity networks are temporal networks representing the close-range proximity among humans in a physical space. They have been extensively studied in the past 15 years as they are critical for understanding the spreading of diseases and information among humans. Here we address the problem of mapping human proximity networks into hyperbolic spaces. Each snapshot of these networks is often very sparse, consisting of a small number of interacting (i.e., non-zero degree) nodes. Yet, we show that the time-aggregated representation of such systems over sufficiently large periods can be meaningfully embedded into the hyperbolic space, using methods developed for traditional (non-mobile) complex networks. We justify this compatibility theoretically and validate it experimentally. We produce hyperbolic maps of six different real systems, and show that the maps can be used to identify communities, facilitate efficient greedy routing on the temporal network, and predict future links with significant precision. Further, we show that epidemic arrival times are positively correlated with the hyperbolic distance from the infection sources in the maps. Thus, hyperbolic embedding could also provide a new perspective for understanding and predicting the behavior of epidemic spreading in human proximity systems.
\end{abstract}

\maketitle

\section{Introduction}

Understanding the time-varying proximity patterns among humans in a physical space is crucial for better understanding the transmission of airborne diseases, the efficiency of information dissemination, social behavior, and influence~\cite{hui_paper, chaintreau_paper, thomas_paper, MITData, FFData, BarratOverview2015, HolmeStructure2016, HolmeVacc2016}. To this end, human proximity networks have been captured in different environments over days, weeks or months~\cite{chaintreau_paper, ConferenceData, PrimaryData, HospitalData, HighSchoolData, OfficeData, MITData, FFData}. Such time-varying networks are represented as a series of static graph snapshots. Each snapshot corresponds to an observation interval or time slot, which typically spans a few seconds to several minutes depending on the devices used to collect the data. The nodes in each snapshot are people and an edge between two nodes signifies that they had been within proximity range during the corresponding slot.  At the finest resolution, each slot spans $20$ s and the proximity range is $1.5$ m. Such networks have been captured by the SocioPatterns collaboration~\cite{SocioPatterns} in closed settings, such as hospitals, schools, scientific conferences and workplaces, and correspond to face-to-face interactions~\cite{HospitalData, PrimaryData, HighSchoolData, ConferenceData, OfficeData}. At a coarser resolution, each snapshot spans several minutes and proximity range can be up to $10$ m or more. Such networks have been captured in university dormitories, residential communities and university campuses~\cite{DartmouthData, MITData, FFData}.

Irrespective of the context, measurement period and measurement method, different human proximity networks have been shown to exhibit similar structural and dynamical properties~\cite{BarratOverview2015, StarniniDevices2017}. Examples of such properties include the broad distributions of contact and intercontact durations~\cite{hui_paper, chaintreau_paper, thomas_paper, StarniniDevices2017}, and the repeated formation of groups that consist of the same people~\cite{SuneRecurrent, flores2018}. Interestingly, these and other properties of human proximity systems can be well reproduced by simple models of mobile interacting agents~\cite{starnini2013, flores2018}. Specifically, in the recently developed force-directed motion model~\cite{flores2018} similarities among agents act as forces that direct the agents' motion toward other agents in the physical space and determine the duration of their interactions. The probability that two nodes are connected in a snapshot generated by the model resembles the connection probability in the popular $\mathbb{S}^1$ model of traditional (non-mobile) complex networks, which is equivalent to random hyperbolic graphs~\cite{Serrano2008, Krioukov2010, Papadopoulos2019}. Based on this observation, the dynamic-$\mathbb{S}^1$ model has been recently suggested as a minimal latent-space model for human proximity networks~\cite{Papadopoulos2019}. The model forgoes the motion component and assumes that each network snapshot is a realization of the $\mathbb{S}^1$ model. The dynamic-$\mathbb{S}^1$ reproduces many of the observed characteristics of human proximity networks, while being mathematically tractable. Several of the model's properties have been proven in Ref.~\cite{Papadopoulos2019}.

Our approach to map human proximity networks into hyperbolic spaces is founded on the dynamic-$\mathbb{S}^1$ model. Specifically, given that the dynamic-$\mathbb{S}^1$ can generate synthetic temporal networks that resemble human proximity networks across a wide range of structural and dynamical characteristics, can we reverse the synthesis and map (embed) human proximity networks into the hyperbolic space, in a way congruent with the model? Would the results of such mapping be meaningful? And could the obtained maps facilitate applications, such as community detection, routing on the temporal network, prediction of future links, and prediction of epidemic arrival times? 

Here we provide the affirmative answers to these questions. Our approach is based on embedding the time-aggregated network of human proximity systems over an adequately large observation period, using methods developed for traditional complex networks that are based on the $\mathbb{S}^1$ model~\cite{GarciaPerez2019}. In the time-aggregated network, two nodes are connected if they are connected in at least one network snapshot during the observation period. We justify this approach theoretically by showing that the connection probability in the time-aggregated network in the dynamic-$\mathbb{S}^1$ model resembles the connection probability in the $\mathbb{S}^1$ model, and explicitly validate it in synthetic networks. Following this approach, we produce hyperbolic maps of six different real systems, and show that the obtained maps are meaningful: they can identify actual node communities, they can facilitate efficient greedy routing on the temporal network, and they can predict future links with significant precision. Further, we show that epidemic arrival times in the temporal network are positively correlated with the hyperbolic distance from the infection sources in the maps.

\section{Results}

\subsection{Data}

We consider the following face-to-face interaction networks from SocioPatterns~\cite{SocioPatterns}. (i) A hospital ward in Lyon~\cite{HospitalData}, which corresponds to interactions involving patients and healthcare workers during five observation days. (ii) A primary school in Lyon~\cite{PrimaryData}, which corresponds to interactions involving children and teachers of ten different classes during two days. (iii) A scientific conference in Turin~\cite{ConferenceData}, which corresponds to interactions among conference attendees during two and a half days. (iv) A high school in Marseilles~\cite{HighSchoolData}, which corresponds to interactions among students of nine different classes during five days. And (v) an office building in Saint Maurice~\cite{Office18Data}, which corresponds to interactions among employees of $12$ different departments during ten days. Each snapshot of these networks corresponds to an observation interval (time slot) of $20$ s, while proximity was recorded if participants were within $1.5$ m in front of each other. 

We also consider the Friends \& Family Bluetooth-based proximity network~\cite{FFData}. This network corresponds to the proximities among residents of a community adjacent to a major research university in the US during several observation months. We consider the data recorded in March~2011. Each snapshot corresponds to an observation interval of 5 min, while proximity was recorded if participants were within a radius of $10$ m from each other. Thus  proximity in this network does not imply face-to-face interaction. Table~\ref{tab:networks} gives an overview of the data.

\begin{table}[htb!]
\centering
\begin{tabular}{|c|c|c|c|c|c|c|c|}
\hline 
Network & Days & $N$ & $\tau$ & $\bar{n}$ & $\bar{k}$ & $\tilde{\bar{k}}$ & $T$ \\ 
\hline 
Hospital & 5 & 75 & 17376 & 2.9 & 0.05 & 30 & 0.84 \\ 
\hline 
Primary school & 2 & 242 & 5846 & 30 & 0.18 & 69 & 0.72 \\ 
\hline 
Conference & 2.5 & 113 & 10618 & 3.3 & 0.03 & 39 & 0.85 \\ 
\hline 
High school & 5 & 327 & 18179 & 17 & 0.06 & 36 & 0.61 \\ 
\hline 
Office building & 10 & 217 & 49678 & 2.8 & 0.01 & 39 & 0.74 \\ 
\hline
Friends \& Family & 31 & 112 & 7317 & 58 & 1.5 & 57 & 0.48 \\
\hline 
\end{tabular}
\caption{Overview of the considered real networks. $N$ is the number of nodes, $\tau$ is the total number of time slots (snapshots), $\bar{n}$ is the average number of interacting (i.e., non-zero degree) nodes per snapshot, $\bar{k}$ is the average node degree per snapshot, $\tilde{\bar{k}}$ is the average degree in the time-aggregated network formed over the full observation duration $\tau$, and parameter $T$ is the network temperature used in the dynamic-$\mathbb{S}^1$ model to generate synthetic counterparts of the real systems (see Appendix~\ref{sec:ds1}). The table also shows the number of observation days for each network.
\label{tab:networks}}
\end{table}

\subsection{$\mathbb{S}^1$ and dynamic-$\mathbb{S}^1$ models}

We first provide an overview of the $\mathbb{S}^1$ and dynamic-$\mathbb{S}^1$ models. In the next section we show that the connection probability in the time-aggregated network in the latter resembles the connection probability in the former. Based on this equivalence, we then map the time-aggregated networks of the considered real data to the hyperbolic space using a recently developed method that is based on the $\mathbb{S}^1$ model.

\subsubsection{$\mathbb{S}^1$ model} 

The $\mathbb{S}^1$ model~\cite{Serrano2008, Krioukov2010} can generate synthetic network snapshots that possess many of the common structural properties of real networks, including heterogeneous or homogeneous degree distributions, strong clustering, and the small-world property. In the model, each node has latent (or hidden) variables $\kappa, \theta$. The latent variable $\kappa$ is proportional to the node's expected degree in the resulting network. The latent variable $\theta$ is the angular similarity coordinate of the node on a circle of radius $R=N/2\pi$, where $N$ is the total number of nodes. To construct a network with the model that has size $N$, average node degree $\bar{k}$, and temperature $T \in (0,1)$, we perform the following steps. First, for each node $i=1, 2, \ldots, N$, we sample its angular coordinate $\theta_i$ uniformly at random from $[0, 2\pi]$, and its degree variable $\kappa_i$ from a probability density function $\rho(\kappa)$. Then, we connect every pair of nodes $i, j$ with the Fermi-Dirac connection probability
\begin{equation}
\label{eq:fermi}
p(\chi_{ij}) = \frac{1}{1+\chi_{ij}^{1/T}}.
\end{equation}
In the last expression, $\chi_{ij}$ is the effective distance between nodes $i$ and $j$,
\begin{equation}
\label{eq:effdist}
\chi_{ij}=\frac{R\Delta\theta_{ij}}{\mu\kappa_i\kappa_j},
\end{equation}
where $\Delta \theta_{ij}=\pi - | \pi -|\theta_i - \theta_j||$ is the similarity distance between nodes $i$ and $j$. Parameter $\mu$ in~(\ref{eq:effdist}) is derived from the condition that the expected degree in the network is indeed $\bar{k}$, yielding 
\begin{equation}
\label{eq:mu}
\mu=\frac{\bar{k}\sin(T\pi)}{2\bar{\kappa}^{2}T\pi},
\end{equation}
where $\bar{\kappa}=\int \kappa\rho(\kappa)d\kappa$. 

The degree distribution $P(k)$ in the resulting network has a similar functional form as $\rho(\kappa)$. Thus, the model can generate networks with \emph{any} degree distribution depending on $\rho(\kappa)$. For instance, a power law degree distribution with exponent $\gamma > 2$ is obtained if $\rho(\kappa) \propto \kappa^{-\gamma}$, while a Poisson degree distribution with mean $\bar{k}$ is obtained if $\rho(\kappa)=\delta(\kappa-\bar{k})$, where $\delta(x)$ is  the Dirac delta function~\cite{Boguna2003, Serrano2008}. Smaller values of the temperature $T$ favor connections at smaller effective distances and increase the average clustering in the network~\cite{Krioukov2010}.  The $\mathbb{S}^1$ model is equivalent to random hyperbolic graphs, i.e., to the hyperbolic $\mathbb{H}^2$ model~\cite{Krioukov2010}, after transforming the degree variables $\kappa_i$ to radial coordinates $r_i$ via
\begin{equation}
\label{eq:radial}
r_i=\hat{R}-2\ln\frac{\kappa_i}{\kappa_0},
\end{equation}
where $\kappa_0$ is the smallest $\kappa_i$ and $\hat{R}=2\ln{[N/(\pi\mu\kappa_0^2)]}$ is the radius of the hyperbolic disk where all nodes reside. After this change of variables, the effective distance in~(\ref{eq:effdist}) becomes $\chi_{ij}=e^{\frac{1}{2}(x_{ij}-\hat{R})}$, where $x_{ij} = r_i+r_j+2\ln{(\Delta \theta_{ij}/2)}$ is approximately the hyperbolic distance between nodes $i$ and $j$~\cite{Krioukov2010}. Therefore, we can refer to the degree variables $\kappa_i$ as ``coordinates'' and use terms \emph{effective distance} and \emph{hyperbolic distance} interchangeably. 

Given the ability of the $\mathbb{S}^1/\mathbb{H}^2$ model to construct synthetic networks that resemble real networks, several methods have been developed to map real networks into the hyperbolic plane, i.e., to infer the nodes' latent coordinates $r$ (or $\kappa$) and $\theta$, according to the model~\cite{Boguna2010, frag:hypermap, frag:hypermap_cn, Alanis2016, blasius16, GarciaPerez2019}. The hyperbolic maps produced by these methods have been shown to be meaningful, and have been efficiently used in applications such as community detection, greedy routing and link prediction~\cite{Boguna2010, Papadopoulos2012, frag:hypermap, frag:hypermap_cn, blasius16, geometry:multilayer, Alanis2016, Ortiz2017, Alanis2018, Allard2020}. Model-free mapping methods have also been developed~\cite{carlo1}. Further, on a related note, there is a large body of work on embedding both static and temporal networks into Euclidean spaces, e.g., see Refs.~\cite{KimSurvey, CuiSurvey, weg2vec}, and references therein. However,  no prior work has considered embedding temporal networks into hyperbolic spaces, which provide a more accurate reflection of the geometry of real networks~\cite{Papadopoulos2012}.

\subsubsection{dynamic-$\mathbb{S}^1$ model}

The dynamic-$\mathbb{S}^1$ model is based on the $\mathbb{S}^1$ model and has been shown to reproduce many of the observed structural and dynamical properties of human proximity networks~\cite{Papadopoulos2019}. The dynamic-$\mathbb{S}^{1}$ models a sequence of network snapshots, $G_t$, $t=1,\ldots, \tau$. Each snapshot is a realization of the $\mathbb{S}^{1}$ model. Therefore, there are $N$ nodes that are assigned latent coordinates $\kappa, \theta$ as in the $\mathbb{S}^{1}$ model, which remain fixed. The temperature $T \in (0,1)$ is also fixed, while each snapshot $G_t$ is allowed to have a different average degree $\bar{k}_t, t=1, \ldots, \tau$. The snapshots are generated according to the following simple rules:
\begin{enumerate}
\item[(1)] at each time step $t=1, \ldots, \tau$, snapshot $G_t$ starts with $N$ disconnected nodes, while $\bar{k}$ in~(\ref{eq:mu}) is set equal to $\bar{k}_t$;
\item[(2)] each pair of nodes $i, j$ connects with probability given by~(\ref{eq:fermi});
\item[(3)] at time $t+1$, all the edges in snapshot $G_t$ are deleted and the process starts over again to generate snapshot $G_{t+1}$.
\end{enumerate}
As shown in Ref.~\cite{Papadopoulos2019}, temperature $T$ plays a central role in network dynamics in the model, dictating the distributions of contact and intercontact durations, the time-aggregated node degrees, and the formation of unique and recurrent components. Specifically, the contact and intercontact distributions are power laws with exponents $2+T$ and $2-T$, respectively. These exponents lie within the ranges observed in real systems~\cite{Papadopoulos2019}. Further, larger values of $T$ increase the connection probability at larger distances, which increases the time-aggregated node degrees. For the same reason, larger values of $T$ increase the number of unique components formed, while decreasing the number of recurrent components. See Ref.~\cite{Papadopoulos2019} for further details.

\subsection{Hyperbolic mapping of human proximity networks}

\subsubsection{Theoretical considerations} 

\begin{figure}[t!]
\centering
\includegraphics[width=\linewidth, height=1.3in]{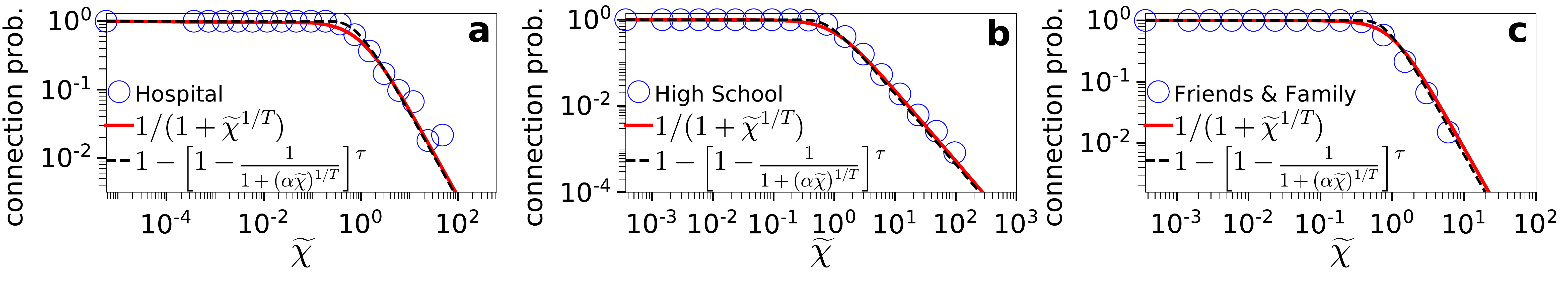}
\caption{Connection probability in the time-aggregated network versus Fermi-Dirac connection probability. The results correspond to the synthetic counterparts of the hospital, high school and Friends \& Family, constructed using the dynamic-$\mathbb{S}^1$ model as described in Appendix~\ref{sec:ds1}. The blue circles show the empirical connection probabilities. The solid red and dashed black lines correspond to~(\ref{eq:fermi}) and (\ref{eq:con_dS1_agg_v2}), respectively. The values of parameters $T$ and $\tau$ in each case are as shown in Table~\ref{tab:networks}, while $\alpha=\tau^T/\Gamma(1+T)$. Similar results hold for the counterparts of the rest of the real systems (see Appendix~\ref{sec:connection_prob_app}).}
\label{fig:fermi_vs_real}
\end{figure}

Assuming that a sequence of network snapshots $G_t$, $t=1,\ldots, \tau$, has been generated by the dynamic-$\mathbb{S}^1$ model, we show below that we can accurately infer the nodes' latent coordinates $\kappa, \theta$ from the time-aggregated network, using existing methods that are based on the $\mathbb{S}^1$ model. This is justified by the fact that the connection probability in the time-aggregated network of the dynamic-$\mathbb{S}^1$ resembles the connection probability in the $\mathbb{S}^1$. Indeed, in the time-aggregated network two nodes are connected if they are connected in at least one of the snapshots. Assuming for simplicity that each snapshot has the same average degree $\bar{k}_t=\bar{k}$, the connection probability in the time-aggregated network of the dynamic-$\mathbb{S}^1$, is
\begin{equation}
\label{eq:con_dS1_agg_v1}
P(\chi_{ij})=1-\left[1-p(\chi_{ij})\right]^\tau,
\end{equation}
where $p(\chi_{ij})$ is given by~(\ref{eq:fermi}). Further, as shown in Ref.~\cite{Papadopoulos2019}, the expected degree of a node in the time-aggregated network, $\tilde{\kappa}$, is related to the node's latent degree $\kappa$, via
\begin{equation}
\label{eq:alpha}
\tilde{\kappa}=\alpha \kappa,
\end{equation}
where $\alpha=\tau^T/\Gamma(1+T)$ for $\tau \gg 1$, and $\Gamma$ is the gamma function. Eq.~(\ref{eq:alpha}) is derived in the thermodynamic limit ($N \to \infty$), where there are no cutoffs imposed to node degrees by the network size. We can therefore rewrite~(\ref{eq:con_dS1_agg_v1}) as
\begin{align}
\label{eq:con_dS1_agg_v2}
P(\tilde{\chi}_{ij})&=1-\left[1-p(\alpha\tilde{\chi}_{ij})\right]^\tau\\
\nonumber &=1-\left\{1+\frac{1}{\tau}\left[\frac{\Gamma(1+T)}{\tilde{\chi}_{ij}}\right]^{1/T}\right\}^{-\tau}\\
\label{eq:exp_approx}
&\approx 1-e^{-\left[\frac{\Gamma(1+T)}{\tilde{\chi}_{ij}}\right]^{1/T}},
\end{align}
where
\begin{equation}
\label{eq:tilde_chi}
\tilde{\chi}_{ij} = \frac {R\Delta\theta_{ij}}{\tilde{\mu}\tilde{\kappa}_i\tilde{\kappa}_j}=\frac{\chi_{ij}}{\alpha}
\end{equation} 
is the effective distance between nodes $i$ and $j$ in the time-aggregated network, while $\tilde{\mu}=\mu/\alpha$. The exponential approximation in~(\ref{eq:exp_approx}) holds for sufficiently large $\tau$. We also note that since $T \in (0,1)$, $0.88 < \Gamma(1+T) < 1$. At large distances, $\tilde{\chi}_{ij} \gg \Gamma(1+T)$, we can use the approximation $e^{-x} \approx 1-x$ in~(\ref{eq:exp_approx}), to write
\begin{equation}
\label{eq:approx1}
P(\tilde{\chi}_{ij}) \approx \frac{C}{\tilde{\chi}_{ij}^{1/T}} \propto \frac{1}{\tilde{\chi}_{ij}^{1/T}} \approx p(\tilde{\chi}_{ij}),
\end{equation}
where $p(x)$ is given by~(\ref{eq:fermi}), while $C=\Gamma{(1+T)}^{1/T}$, $0.56 < C < 1$. At small distances, $\tilde{\chi}_{ij} \ll  \Gamma(1+T)$, the exponential in~(\ref{eq:exp_approx}) is much smaller than one, and we can write $P(\tilde{\chi}_{ij}) \approx 1 \approx p(\tilde{\chi}_{ij})$. In other words, at both small and large effective distances $\tilde{\chi}_{ij}$, the connection probability in the time-aggregated network resembles the Fermi-Dirac connection probability in the $\mathbb{S}^1$ model. Fig.~\ref{fig:fermi_vs_real} illustrates this effect in the time-aggregated networks of synthetic counterparts of real systems, whose snapshots can also have different average degrees $\bar{k}_t, t=1, \ldots, \tau$ (see Appendix~\ref{sec:ds1}). 

\begin{figure}[b!]
\centering
\includegraphics[width=\linewidth, height=2.3in]{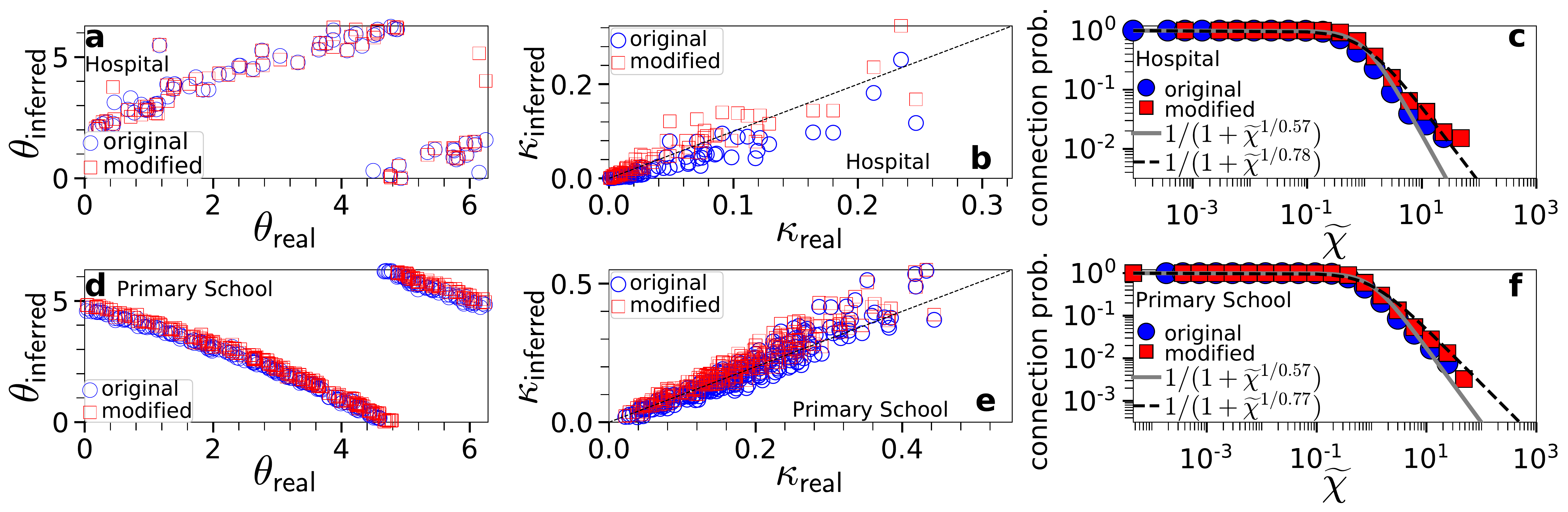}
\caption{Inference of latent coordinates $(\kappa, \theta)$ with the original and modified versions of Mercator. The top row corresponds to a synthetic counterpart of the hospital, while the bottom row to a synthetic counterpart of the primary school. Both versions of Mercator are applied to the corresponding time-aggregated network formed over the full duration $\tau$ in Table~\ref{tab:networks}. (\textbf{a} and \textbf{d}) Inferred versus real $\theta$. (\textbf{b} and \textbf{e}) Inferred versus real $\kappa$. For each node, $\kappa_{\textnormal{inferred}}$ is estimated as $\kappa_{\textnormal{inferred}}=\tilde{\kappa}/\alpha$, where $\tilde{\kappa}$ is the node's inferred latent degree in the time-aggregated network, while $\alpha=\tau^T/\Gamma(1+T)$, with $\tau$ as in Table~\ref{tab:networks} and $T$ as inferred by each version of Mercator. (\textbf{c} and \textbf{f}) Connection probability as a function of the effective distance $\tilde{\chi}$ in the time-aggregated network computed using the inferred coordinates $(\tilde{\kappa}, \theta)$. The solid grey and dashed black lines correspond to~(\ref{eq:fermi}) with temperature $T$ as inferred by each version of Mercator. For the two networks, the original version estimates $T=0.57$, the modified version estimates $T=0.78$ and $0.77$, while the actual values are $T=0.84$ and $0.72$. In general, the modified version estimates values of $T$ closer to the actual values. However, both versions of Mercator perform remarkably well at estimating the nodes' latent coordinates ($\kappa, \theta$). We note that due to rotational symmetry of the model, the inferred angles can be globally shifted compared to the real angles by any value in $[0, 2\pi]$.
\label{fig:inf_mercator}}
\end{figure}

Given this equivalence, in Fig.~\ref{fig:inf_mercator} we apply Mercator, a recently developed embedding method based on the $\mathbb{S}^1$ model~\cite{GarciaPerez2019}, to the time-aggregated network of the synthetic counterparts of the hospital and primary school. Mercator infers the nodes' coordinates $(\tilde{\kappa}, \theta)$ from the time-aggregated network (see Appendix~\ref{sec:mercator}), and from $\tilde{\kappa}$ we estimate $\kappa$ using~(\ref{eq:alpha}). We also modified Mercator to use the connection probability in~(\ref{eq:con_dS1_agg_v2}) instead of the connection probability in~(\ref{eq:fermi}) (see Appendix~\ref{sec:mod_mercator_app}). Fig.~\ref{fig:inf_mercator} shows that the two versions of Mercator perform similarly, inferring the nodes' latent coordinates remarkably well. Similar results hold for the synthetic counterparts of the rest of the real systems (Appendix~\ref{sec:inference_app}). In the rest of the paper, we use the original version of Mercator as its implementation is simpler and does not require knowledge of parameter $\tau$. 

\subsubsection{Aggregation interval}  

\begin{figure}[h!]
\centering
\includegraphics[width=\linewidth, height=1.1in]{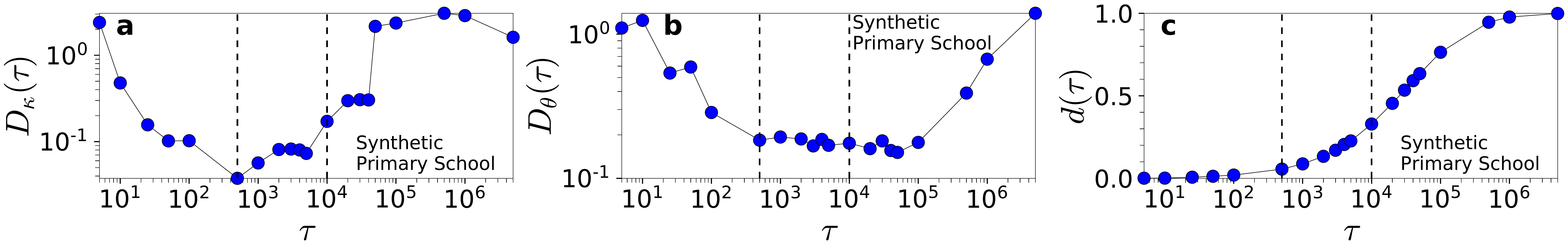}
\caption{Inference accuracy~vs.~aggregation interval. The results correspond to a synthetic counterpart of the primary school constructed using the dynamic-$\mathbb{S}^1$ model. \textbf{(a)} Average difference between the inferred and real latent degrees as a function of the aggregation interval $\tau$, $D_\kappa(\tau)=\sum_{i=1}^{N}|\kappa_{\textnormal{inferred}}^i-\kappa_{\textnormal{real}}^i|/N$, where $\kappa_{\textnormal{inferred}}^i$ ($\kappa_{\textnormal{real}}^i$) is the inferred (real) latent degree of node $i$. \textbf{(b)}~Same as in (a) but for the average difference between the inferred and real angular coordinates, $D_\theta(\tau)=\sum_{i=1}^{N}|\theta_{\textnormal{inferred}}^i-\theta_{\textnormal{real}}^i|/N$. Before computing $D_\theta(\tau)$, the inferred angles are globally shifted such that the sum of the squared distances between real and inferred angles is minimized (to this end, we apply a Procrustean rotation~\cite{procrustes_ref}, see Appendix~\ref{sec:aggr_and_rot} for details). \textbf{(c)} Density of the time-aggregated network as a function of $\tau$, $d(\tau)=2 L/[N(N-1)]$, where $L$ is the number of links in the network. The vertical dashed lines indicate the interval $500 \leq \tau \leq 10000$. In this interval, $D_\kappa(\tau) < 0.2 $, $D_\theta(\tau) <  0.2 $, and $0.06 < d(\tau) < 0.33$.
\label{fig:aggregation_effect}}
\end{figure}

As the aggregation interval $\tau$ increases, the time-aggregated network becomes denser, eventually turning into a fully connected network. This can be seen in~(\ref{eq:con_dS1_agg_v1}), where irrespective of network size, at $\tau \to \infty$, $P(\chi_{ij}) \to 1, \forall i, j$. Further, at $\tau \to \infty$, $\alpha \to \infty$, and by~(\ref{eq:tilde_chi}) $\tilde{\chi}_{ij} \to 0,~\forall i, j$. Clearly, no meaningful inference can be made in a fully connected network as all nodes ``look the same''. Thus for an accurate inference of the nodes' coordinates the interval $\tau$ has to be sufficiently small such that the corresponding time-aggregated network is not too dense. On the other hand, for intervals $\tau$ that are not sufficiently large there may not be enough data to allow accurate inference, as network snapshots are often very sparse in human proximity systems, consisting of only a fraction of nodes (Table~\ref{tab:networks}). This effect is illustrated in Fig.~\ref{fig:aggregation_effect}, where we quantify the difference between real and inferred coordinates as a function of $\tau$ in a synthetic counterpart of the primary school. We see in Fig.~\ref{fig:aggregation_effect} that there is a wide range of adequately large $\tau$ values, e.g., $500 < \tau < 10000$, where the accuracy of inference for both $\kappa$ and $\theta$ is simultaneously high, while as $\tau$ becomes too large or too small accuracy deteriorates. Similar results hold for the counterparts of the rest of the considered real systems (Appendix~\ref{sec:aggr_and_rot}). The exact range of $\tau$ values where inference accuracy is high depends on the system's parameters, e.g., sparser networks (lower average snapshot degree) allow aggregation over longer intervals, as it takes longer for the time-aggregated network to become too dense. Further, our results with the synthetic counterparts suggest that daily aggregation intervals should be sufficient for accurate inference in most cases. Indeed, in this work we embed the time-aggregated networks of the considered real systems formed over the full observation durations $\tau$ in Table~\ref{tab:networks}, as well as corresponding time-aggregated networks formed over individual observation days, obtaining in both cases meaningful results.

\subsubsection{Hyperbolic maps of real systems} 

\begin{figure}[t!]
\centering
\includegraphics[width=\linewidth, height=5.1in]{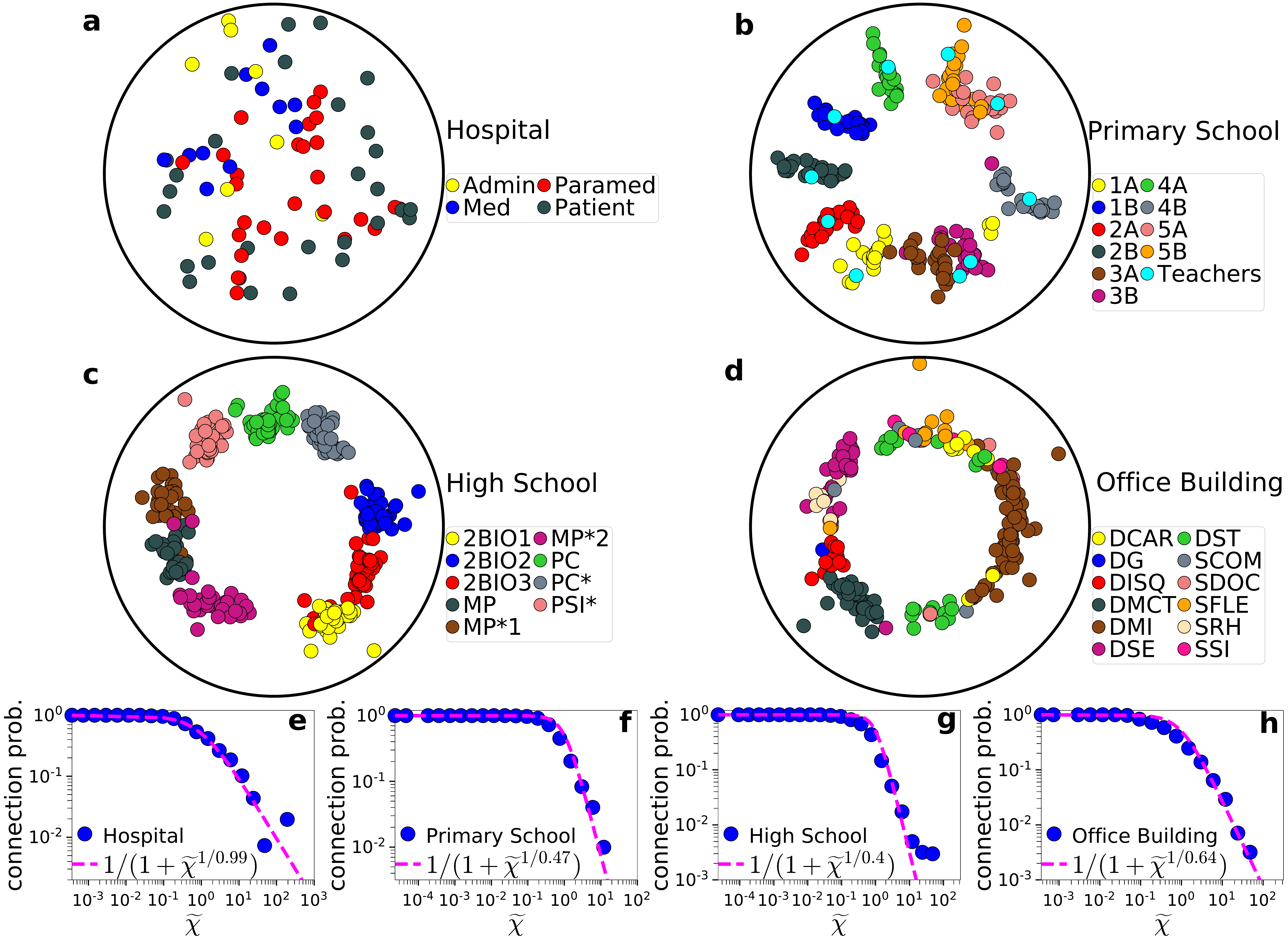}
\caption{Hyperbolic embeddings of human proximity networks. (\textbf{a}-\textbf{d}) Hyperbolic maps of the time-aggregated networks of the hospital, primary school, high school and office building. In each case we consider the time-aggregated network formed over the full observation duration $\tau$ shown in Table~\ref{tab:networks}. The nodes are positioned according to their inferred hyperbolic coordinates ($r, \theta$) in the time-aggregated network [the radial coordinates $r$ are computed using~(\ref{eq:radial})]. The nodes are colored according to group membership information available in the metadata of each network. In the hospital, the nodes are administrative staff (Admin), medical doctors (Med), nurses and nurses' aides (Paramed), and patients (Patient). In the primary school, the nodes are teachers and students of the following classes: 1st grade (1A, 1B), 2nd grade (2A, 2B), 3rd grade (3A, 3B), 4th grade (4A, 4B), and 5th grade (5A, 5B). In the high school, the nodes are students of nine different classes with the following specializations: biology (2BIO1, 2BIO2, 2BIO3), mathematics and physics (MP, MP*1, MP*2), physics and chemistry (PC, PC*), and engineering studies (PSI*). In the office building, the nodes are employees working in different departments such as scientific direction (DISQ), chronic diseases and traumatisms (DMCT), department of health and environment (DSE), human resources (SRH), and logistics (SFLE). (\textbf{e}-\textbf{h}) Corresponding empirical connection probabilities as a function of the effective distance $\tilde{\chi}$. The pink dashed lines correspond to~(\ref{eq:fermi}) with temperatures $T$ as inferred by Mercator, $T=0.99$, $0.47$, $0.40$ and $0.64$, respectively. The maps for the conference and Friends \& Family can be found in Appendix~\ref{sec:maps_app}. Daily hyperbolic maps for each real system can be found in Appendix~\ref{sec:stability_app}.}
\label{fig:hypermaps}
\end{figure}

In Fig.~\ref{fig:hypermaps} we apply Mercator to the time-aggregated network of the real networks in Table~\ref{tab:networks} and visualize the obtained hyperbolic maps and the corresponding connection probabilities. We see that the embeddings are meaningful, as we can identify in them actual node communities that correspond to groups of nodes located close to each other in the angular similarity space. These communities reflect the organization of students and teachers into classes (Figs.~\ref{fig:hypermaps}b and~\ref{fig:hypermaps}c), employees into departments (Fig.~\ref{fig:hypermaps}d), while no communities can be identified in the hospital (Fig.~\ref{fig:hypermaps}a). In all cases, we see a good match between empirical and theoretical connection probabilities (Figs.~\ref{fig:hypermaps}e-h). Next, we turn our attention to greedy routing.

\subsection{Human-to-human greedy routing}

A problem of significant interest in mobile networking is how to efficiently route data in opportunistic networks, like human proximity systems, where the mobility of nodes creates contact opportunities among nodes that can be used to connect parts of the network that are otherwise disconnected~\cite{hui_paper, chaintreau_paper, thomas_paper, ContiOppurtunisticOverview}. Motivated by this problem, and by the remarkable efficiency of hyperbolic greedy routing in traditional complex networks~\cite{Boguna2010, Ortiz2017, Allard2020}, we investigate here if hyperbolic greedy routing can facilitate navigation in human proximity systems. To this end, we consider the following simplest greedy routing process, which performs routing on the temporal network using the coordinates inferred from the time-aggregated network.

\emph{Human-to-human greedy routing (H2H-GR)}. In H2H-GR, a node's address is its coordinates $(\tilde{\kappa}, \theta)$, and each node knows its own address, the addresses of its neighbors (nodes currently within proximity range), and the destination address written in the packet. A node holding the packet (carrier) forwards the packet to its neighbor with the smallest effective distance to the destination, but only if that distance is smaller than the distance between the carrier and the destination. Otherwise, or if the carrier currently has no neighbors, the carrier keeps the packet.  Clearly, a carrier delivers the packet to the destination if the latter is its neighbor. We note that there are no routing loops in H2H-GR, i.e., no node receives the same packet twice. Indeed, consider for instance a packet from a node $i_0$ to a node $i_n$, which has followed the path $\{i_0, i_1, i_2, \ldots, i_{n-1}, i_n\}$. This means that $\tilde{\chi}_{i_0 i_n} > \tilde{\chi}_{i_1 i_n} > \tilde{\chi}_{i_2 i_n} > \ldots > \tilde{\chi}_{i_{n-1} i_n}$, where $\tilde{\chi}_{i_k i_n}$ is the effective distance between nodes $i_k$ and $i_n$. A node $i_k$ in the path never forwards the packet to a node $i_l$ with $l < k$, i.e., to a node that has seen the packet before, because $\tilde{\chi}_{i_l i_n} > \tilde{\chi}_{i_k i_n}$. We also note that in the thermodynamic limit ($N \to \infty$), there is a non-zero probability that a packet constantly moves closer to the destination but never actually reaches it. This event could theoretically occur at $N \to \infty$, as there could be a countably infinite number of intermediate nodes with asymptotically closer effective distances to the destination. In reality such event can never occur since the number of nodes $N$ is bounded.

For each network in Table~\ref{tab:networks}, we simulate H2H-GR in one of its observation days. We consider the following two cases: i) H2H-GR that uses the nodes' coordinates inferred from the time-aggregated network of the considered day (current coordinates); and ii) H2H-GR that uses the nodes' coordinates inferred from the time-aggregated network of the previous day (previous coordinates). In the time-aggregated network of a day, two nodes are connected if they are connected in at least one network snapshot in the day. We compare these two cases to a baseline \emph{random routing strategy (H2H-RR)}, where the carrier first determines the set of its neighbors that have never received the packet before, and then forwards the packet to one of these neighbors at random. If the destination is a neighbor the carrier forwards the packet to it. The carrier keeps the packet if it currently has no neighbors, or if all of its neighbors have received the packet before. Thus, there are no routing loops in H2H-RR either.

\emph{Performance metrics}. We evaluate the performance of the algorithms according to the following two metrics: i)~the percentage of successful paths, $p_s$, which is the proportion of paths that reach their destinations by the end of the considered day; and ii)~the average stretch over the successful paths, $\bar{s}$. We define the stretch as the ratio of the hop-lengths of the paths found by the algorithms to the corresponding shortest time-respecting paths~\cite{Holme2019} in the network.

The results are shown in Table~\ref{tab:gr_real}. We see that H2H-GR that uses the current coordinates significantly outperforms H2H-RR in both success ratio and stretch. The improvement can be quite significant. For instance, in the primary school the success ratio increases from $34$\% to $82$\%, while the average stretch decreases from $24.9$ to $3.9$. Similarly, in the hospital the success ratio increases from $38$\% to $80$\%, while the average stretch decreases from $7$ to $2.2$. These results show that hyperbolic greedy routing can significantly improve navigation. However, the success ratio decreases considerably if H2H-GR uses the previous coordinates. This suggests that the node coordinates change to a considerable extend from one day to the next.  In Appendix~\ref{sec:stability_app}, we verify that this is indeed the case. Nevertheless, H2H-GR that uses the previous coordinates still outperforms H2H-RR with respect to success ratio, while achieving significantly lower stretch similar to the stretch with the current coordinates (Table~\ref{tab:gr_real}). 

\begin{table}[h!]
\centering
\begin{tabular}{|c|c|c|c|}
\hline
Real network & H2H-GR (current coordinates) & H2H-GR (previous coordinates) & H2H-RR \\
\hline
Hospital & $p_s=0.80$, $\bar{s}=2.2$ & $p_s=0.47$, $\bar{s}=2.0$ & $p_s=0.38$, $\bar{s}=7.0$ \\
\hline 
Primary school &  $p_s=0.82$, $\bar{s}=3.9$ & $p_s=0.65$, $\bar{s}=3.6$ & $p_s=0.34$, $\bar{s}=24.9$ \\
\hline
Conference & $p_s=0.70$, $\bar{s}=2.2$ & $p_s=0.35$, $\bar{s}=2.0$ & $p_s=0.29$, $\bar{s}=7.9$ \\
\hline
High school & $p_s=0.29$, $\bar{s}=2.0$ & $p_s=0.13$, $\bar{s}=1.9$ & $p_s=0.07$, $\bar{s}=5.9$ \\
\hline
Office building & $p_s=0.15$, $\bar{s}=1.4$ & $p_s=0.10$, $\bar{s}=1.4$ & $p_s=0.06$, $\bar{s}=2.5$ \\
\hline
Friends \& Family &  $p_s=0.45$, $\bar{s}=1.8$ & $p_s=0.31$, $\bar{s}=2.0$ & $p_s=0.21$, $\bar{s}=5.3$ \\
\hline
\end{tabular}
\caption{
\label{tab:gr_real} 
Success ratio $p_s$ and average stretch $\bar{s}$ of H2H-GR and H2H-RR in real networks. H2H-GR uses the coordinates inferred either from the time-aggregated network of the considered day where routing is performed (current coordinates); or from the time-aggregated network of the previous day (previous coordinates). The considered days in the hospital, primary school, conference, high school and office building are observation days $5, 2, 3, 5$ and $10$, respectively. In Friends \& Family, the considered day is the $31^{\textnormal{st}}$ of March 2011. For a fair comparison with H2H-GR that uses the previous coordinates, we ignore during all routing processes the nodes that exist in the considered day but not in the previous day, since for such nodes we cannot infer their coordinates from the previous day. The percentage of such nodes is $17\%, 3\%, 7\%, 6\%, 14\%$ and $3\%$ for the hospital, primary school, conference, high school, office building and Friends \& Family, respectively. In all cases, routing is performed among all possible source-destination pairs in the considered day that also exist in the previous day.}
\end{table}

Table~\ref{tab:gr_synth} shows the same results for the synthetic counterparts of the real systems, where we can make qualitatively similar observations. Further, we see that H2H-GR achieves higher success ratios using the inferred coordinates in the counterparts compared to the real systems. This is not surprising as the counterparts are by construction maximally congruent with the assumed geometric model (dynamic-$\mathbb{S}^1$). Also, H2H-GR that uses the previous coordinates maintains high success ratios in the counterparts. This is expected, as the coordinates in the counterparts do not change over time. Thus the coordinates inferred from the time-aggregated network of the previous day are quite similar (but not exactly the same) to the ones inferred from the time-aggregated network of the day where routing is performed (see Appendix~\ref{sec:stability_app}).

\begin{table}[t!]
\centering
\begin{tabular}{|c|c|c|c|}
\hline
Synthetic network & H2H-GR (current coordinates) & H2H-GR (previous coordinates) & H2H-RR \\
\hline
Hospital & $p_s=0.92$, $\bar{s}=2.2$ & $p_s=0.78$, $\bar{s}=2.2$ & $p_s=0.42$, $\bar{s}=9.2$ \\
\hline 
Primary school & $p_s=0.98$, $\bar{s}=3.7$ & $p_s=0.97$, $\bar{s}=3.8$ & $p_s=0.53$, $\bar{s}=33.9$ \\
\hline
Conference & $p_s=0.85$, $\bar{s}=2.4$ & $p_s=0.70$, $\bar{s}=2.4$ & $p_s=0.31$, $\bar{s}=9.8$ \\
\hline
High school &  $p_s=0.72$, $\bar{s}=2.7$ & $p_s=0.59$, $\bar{s}=2.4$ & $p_s=0.11$, $\bar{s}=7.8$ \\
\hline
Office building & $p_s=0.26$, $\bar{s}=1.5$ & $p_s=0.17$, $\bar{s}=1.5$ & $p_s=0.06$, $\bar{s}=3.0$ \\
\hline
Friends \& Family & $p_s=0.82$, $\bar{s}=2.2$ & $p_s=0.70$, $\bar{s}=2.3$ & $p_s=0.23$, $\bar{s}=5.4$ \\
\hline
\end{tabular}
\caption{
\label{tab:gr_synth}
Same as in Table~\ref{tab:gr_real} but for the synthetic counterparts of the real systems constructed with the dynamic-$\mathbb{S}^1$ model. 
The results in each case correspond to one temporal network realization, while H2H-GR uses inferred coordinates as in Table~\ref{tab:gr_real}.}
\end{table}

\begin{figure}[b!]
\centering
\includegraphics[width=\linewidth, height=2.5in]{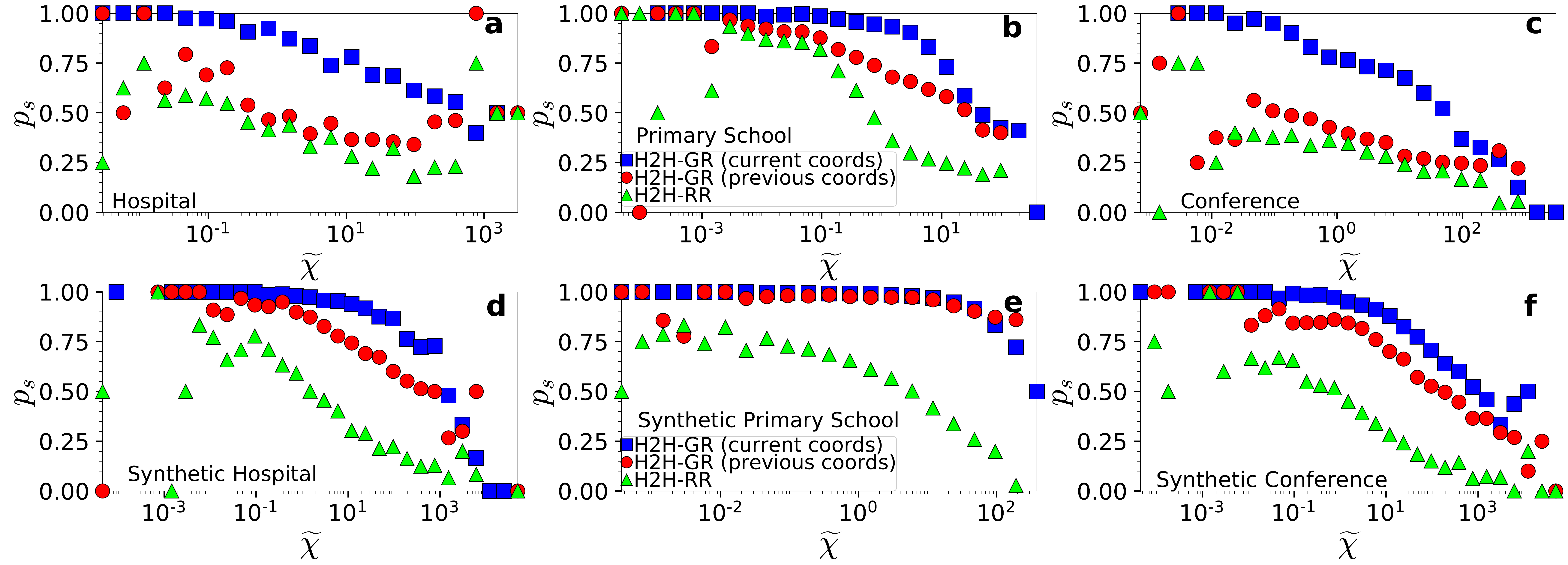}
\caption{Success ratio $p_s$ of H2H-GR and H2H-RR as a function of the effective distance $\tilde{\chi}$ between source-destination pairs. The top row corresponds to the results of the hospital, primary school and conference in Table~\ref{tab:gr_real}, while the bottom row to the results of their synthetic counterparts in Table~\ref{tab:gr_synth}. The success ratio for H2H-RR and H2H-GR that uses the previous coordinates is shown as a function of the effective distance between the pairs in the previous day. Similar results hold for the other real networks and their synthetic counterparts (Appendix~\ref{sec:h2h_gr_app}).}
\label{fig:gr_success}
\end{figure}

\begin{figure}[t!]
\centering
\includegraphics[width=\linewidth, height=2.5in]{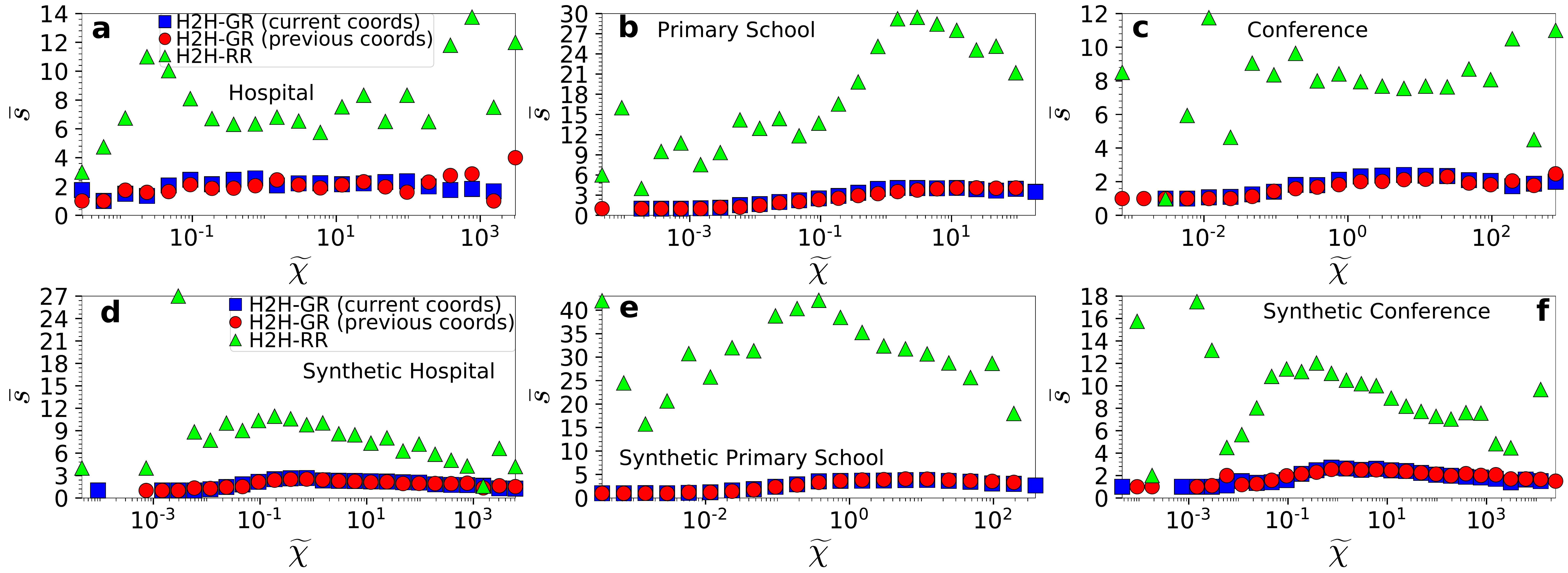}
\caption{Same as in Fig.~\ref{fig:gr_success} but for the average stretch $\bar{s}$. Similar results hold for the other real networks and their synthetic counterparts (Appendix~\ref{sec:h2h_gr_app}).}
\label{fig:gr_stretch}
\end{figure}

The metrics in Tables~\ref{tab:gr_real} and~\ref{tab:gr_synth} are computed across all source-destination pairs.  In Figs.~\ref{fig:gr_success} and~\ref{fig:gr_stretch} we also compute these metrics as a function of the effective distance between the source-destination pairs. We see that H2H-GR that uses the current coordinates achieves high success ratios, approaching $100$\%, as the effective distance between the pairs decreases. As the effective distance between the pairs increases, the success ratio decreases. The average stretch for successful H2H-GR paths is always low.

H2H-RR also achieves considerably high success ratios for pairs separated by small distances (Fig.~\ref{fig:gr_success}). This is because, even though packets in H2H-RR are forwarded to neighbors at random, the neighbors are not random nodes but nodes closer to the carriers in the hyperbolic space. Thus, packets between pairs separated by smaller distances have higher chances of finding their destinations.  However, the stretch of successful paths in H2H-RR is quite high (Fig.~\ref{fig:gr_stretch}).  Further, we see that in real networks the success ratio of H2H-GR that uses the previous coordinates resembles in most cases the one of H2H-RR (Figs.~\ref{fig:gr_success}a-c and Figs.~\ref{fig:gr_si_success}a-c). However, the stretch in H2H-GR is always significantly lower than in H2H-RR (Figs.~\ref{fig:gr_stretch}a-c and Figs.~\ref{fig:gr_si_stretch}a-c).

Taken altogether, these results show that hyperbolic greedy routing can facilitate efficient navigation in human proximity networks. The success ratio for pairs separated by large effective distances can be low (Fig.~\ref{fig:gr_success}). However, it is possible that more sophisticated algorithms than the one considered here could improve the success ratio for such pairs without significantly sacrificing stretch. Further, using coordinates from past embeddings decreases the success ratio. Even though the average stretch remains low, this observation suggests that the evolution of the nodes' coordinates should also be taken into account. Such investigations are beyond the scope of this paper. Finally, we note that in Appendix~\ref{sec:h2h_gr_app}, we consider H2H-GR that uses only the angular similarity distances among the nodes, and find that it performs worse than H2H-GR that uses the effective distances. This means that in addition to node similarities, node expected degrees (or popularities~\cite{Papadopoulos2012}) also matter in H2H-GR, even though the distribution of node degrees in human proximity systems is quite homogeneous~\cite{Papadopoulos2019}.

\subsection{Link prediction}

In this section, we turn our attention to link prediction. We want to see how well we can predict if two nodes are connected in the time-aggregated network of a day, if we know the effective distances among the nodes in the previous day. To this end, for each pair of nodes $i, j$ in the previous day that is also present in the day of interest, we assign a score $s_{ij}=1/\tilde{\chi}_{ij}$, where $\tilde{\chi}_{ij}$ is the inferred effective distance between $i$ and $j$ in the time-aggregated network of the previous day. The higher the $s_{ij}$, the higher is the likelihood that $i$ and $j$ are connected in the day of interest. We call this approach \emph{geometric}. To quantify the quality of link prediction, we use two standard metrics: (i) the \emph{Area Under the Receiver Operating Characteristic curve (AUROC)}; and (ii) the \emph{Area Under the Precision-Recall curve (AUPR)}~\cite{Saito2016}. These metrics are described below.

The AUROC represents the probability that a randomly selected connected pair of nodes is given a higher score than a randomly selected disconnected pair of nodes in the day of interest. The degree to which the AUROC exceeds $0.5$ indicates how much better the method performs than pure chance. As the name suggests, the AUROC is equal to the total area under the \emph{Receiver Operating Characteristic (ROC)} curve. To compute the ROC curve, we order the pairs of nodes in the descending order of their scores, from the largest $s_{ij}$ to the smallest $s_{ij}$, and consider each score to be a threshold. Then, for each threshold we calculate the fraction of connected pairs that are above the threshold (i.e., the True Positive Rate TPR) and the fraction of disconnected pairs that are above the threshold (i.e., the False Positive Rate FPR). Each point on the ROC curve gives the TPR and FPR for the corresponding threshold. When representing the TPR in front of the FPR, a totally random guess would result in a straight line along the diagonal $y = x$, while the degree by which the ROC curve lies above the diagonal indicates how much better the algorithm performs than pure chance. $\textnormal{AUROC} = 1$ means a perfect classification (ordering) of the pairs, where the connected pairs are placed in the top of the ordered list.

The AUPR represents how accurately the method can classify pairs of nodes as connected and disconnected based on their scores. It is equal to the total area under the \emph{Precision-Recall (PR)} curve. To compute the PR curve, we again order the pairs of nodes in the descending order of their scores, and consider each score to be a threshold. Then, for each threshold we calculate the TPR, which is called Recall, and the Precision, which is the fraction of pairs above the threshold that are connected. Each point on the PR curve gives the Precision and Recall for the corresponding threshold. A random guess corresponds to a straight line parallel to the Recall axis at the level where Precision equals the ratio of the number of connected pairs to the total number of pairs. The higher the AUPR the better the method is, while a perfect classifier yields $\textnormal{AUPR}=1$. 

\begin{figure}[t!]
\centering
\includegraphics[width=\linewidth, height=3.8in]{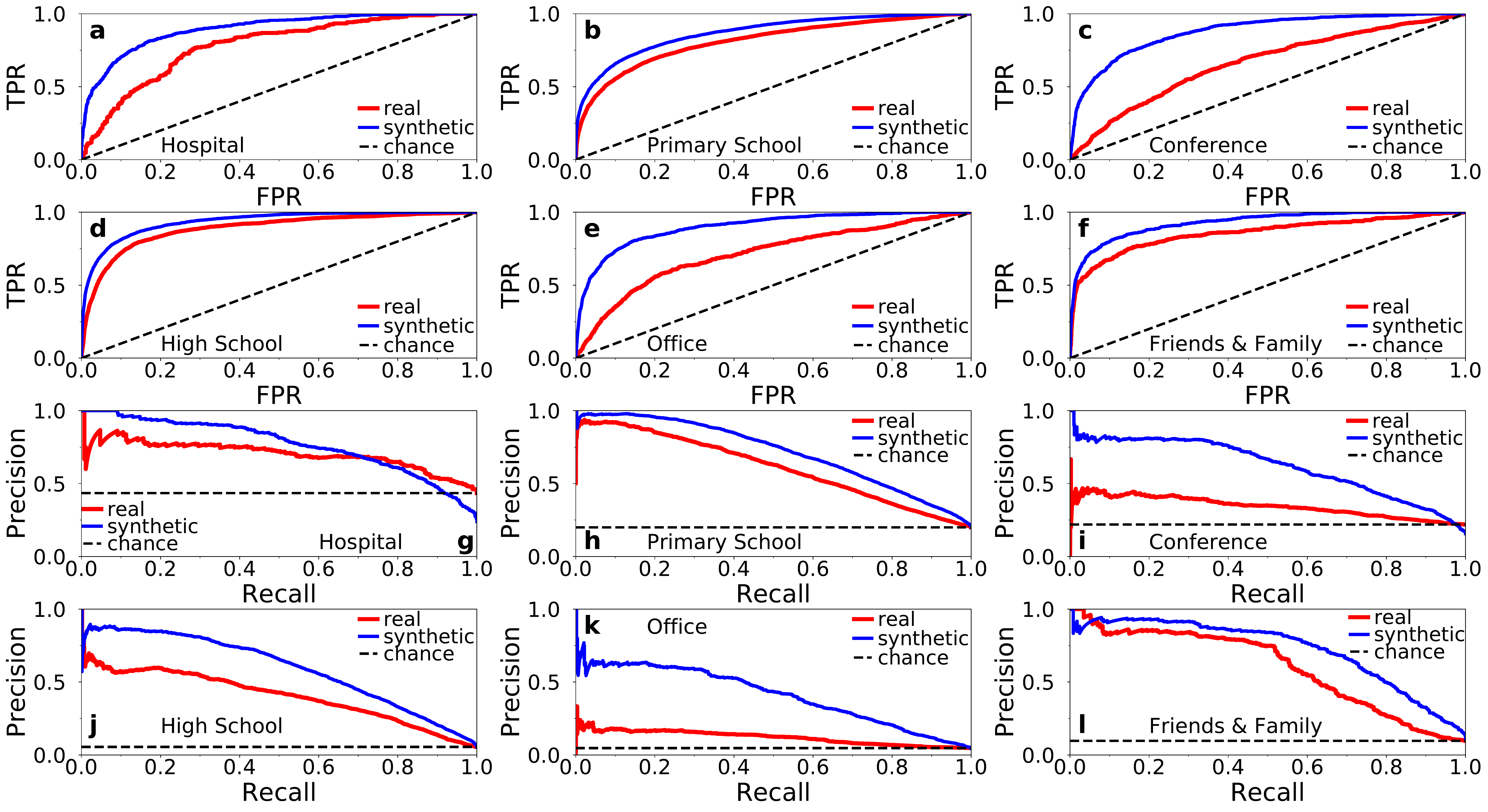}
\caption{ROC and PR curves for geometric link prediction in real networks and their synthetic counterparts. (\textbf{a}-\textbf{f}) show the ROC curves, while (\textbf{g}-\textbf{l}) the PR curves, corresponding to the results in Table~\ref{tab:lp}. The dashed black lines correspond to link prediction based on chance; these lines in (\textbf{g}-\textbf{l}) correspond to the AUPR chance values in Table~\ref{tab:lp}.
\label{fig:lp}}
\end{figure}

The results for the considered real networks and their synthetic counterparts are shown in Table~\ref{tab:lp}. The corresponding ROC and PR curves are shown in Fig.~\ref{fig:lp}. We see that geometric link prediction significantly outperforms chance in all cases. These results constitute another validation that the embeddings are meaningful, and illustrate that they have significant predictive power. As can be seen in Table~\ref{tab:lp} and Fig.~\ref{fig:lp}, link prediction is more accurate in the synthetic counterparts. This is again expected since the counterparts are by construction maximally congruent with the underlying geometric space, while the node coordinates in them do not change over time.  

We also compute the same metrics as in Table~\ref{tab:lp} but for a simple heuristic, where the score $s_{i j}$ between two nodes $i$ and $j$ is the number of common neighbors they have in the time-aggregated network of the previous day (CN approach). The results are shown in Table~\ref{tab:lp_cn}. Interestingly, we see that the performance of the geometric and CN approaches is quite similar in real networks, suggesting that the latter is a good heuristic for link prediction in human proximity systems. The performance of the two approaches is also positively correlated in the synthetic counterparts (Tables~\ref{tab:lp} and~\ref{tab:lp_cn}). This is expected since the smaller the effective distance between two nodes the larger is the \emph{expected} number of common neighbors the nodes have. However, as can be seen in Tables~\ref{tab:lp} and~\ref{tab:lp_cn}, in the counterparts the geometric approach performs better than the CN approach. This suggests that the performance of the former could be further improved in real systems, if more accurate predictions of the node coordinates in the period of interest could be made.

\begin{table}[t!]
\centering
\begin{tabular}{|c|c|c|c|c|c|c|}
\hline
Network & AUROC real & AUPR real &AUROC chance & AUPR chance & AUROC synthetic & AUPR synthetic \\
\hline
Hospital & 0.78 & 0.70 & 0.5 & 0.43 & 0.90 & 0.77 \\
\hline 
Primary school & 0.81 & 0.62 & 0.5 & 0.20 & 0.87 & 0.71 \\
\hline
Conference & 0.66 & 0.34 & 0.5 & 0.22 & 0.88 & 0.62 \\
\hline
High school & 0.89 & 0.40 & 0.5 & 0.05 & 0.94 & 0.59 \\
\hline
Office building & 0.71 & 0.12 & 0.5 & 0.05 & 0.90 & 0.41 \\
\hline
Friends \& Family & 0.86 & 0.60 & 0.5 & 0.10 & 0.93 & 0.72 \\
\hline
\end{tabular}
\caption{AUROC and AUPR for geometric link prediction in real networks and their synthetic counterparts. The day of interest is day $3$ in the hospital and day $2$ in the rest of the networks. Geometric link prediction uses the effective distances among the nodes inferred from the time-aggregated network of the previous day. ``AUPR chance'' corresponds to link prediction based on pure chance in the real networks. It equals the ratio of the number of connected pairs to the total number of pairs in the time-aggregated network of the day of interest. AUPR chance values for the synthetic counterparts are similar as in the real networks and not shown for brevity.
\label{tab:lp}}
\end{table}

\begin{table}[hbt!]
\centering
\begin{tabular}{|c|c|c|c|c|}
\hline
Network & AUROC real & AUPR real & AUROC synthetic & AUPR synthetic \\
\hline
Hospital & 0.75 & 0.79 &  0.85 & 0.69 \\
\hline 
Primary school & 0.79 & 0.52 &  0.84 & 0.62 \\
\hline
Conference & 0.67 & 0.37 &  0.85 & 0.57 \\
\hline
High school & 0.88 & 0.44 &  0.89 & 0.52 \\
\hline
Office building & 0.73 & 0.10 & 0.86 & 0.35 \\
\hline
Friends \& Family & 0.85 & 0.54 & 0.89 & 0.64 \\
\hline
\end{tabular}
\caption{Same as in Table~\ref{tab:lp} but for the CN approach.
\label{tab:lp_cn}}
\end{table}

\subsection{Epidemic spreading}
\label{sec:spreading}

Finally, we consider epidemic spreading. Here, predicting the arrival time of an epidemic is crucial for developing better containment measures for infectious diseases~\cite{gauvin2013,Brockmann2013}. In the context of the global air transportation network, Brockmann and Helbing showed that the epidemic arrival time in a country can be well predicted by the effective distance between the country and the infection source country~\cite{Brockmann2013}. The effective distance between two countries is defined as the length of the shortest weighted path connecting the two countries in the air transportation network, where the weight of a link is a decreasing function of the air traffic between the endpoints of the link~\cite{Brockmann2013}.

In a similar vein, here we show that in human proximity networks, the epidemic arrival time, i.e., the time slot at which a node becomes infected, is positively correlated with the hyperbolic distance between the node and the infected source node in the time-aggregated network. [We note that while in Ref.~\cite{Brockmann2013} the effective distances are directly defined by observable (weighted) path lengths, the effective distances in our case are defined by the nodes' latent coordinates that manifest themselves indirectly via the nodes' connections and disconnections in the (unweighted) time-aggregated network.] To this end, we consider the Susceptible-Infected (SI) epidemic spreading model~\cite{Kermack1927}. In the SI, each node can be in one of two states, susceptible (S) or infected (I). At any time slot infected nodes infect susceptible nodes with whom they are within proximity range, with probability $\beta$. Thus, the transition of states is S$\rightarrow$I. To simulate the SI process on temporal networks we use the dynamic SI implementation of the Network Diffusion library~\cite{NDlib}. 

\begin{figure}[t!]
\centering
\includegraphics[width=\linewidth]{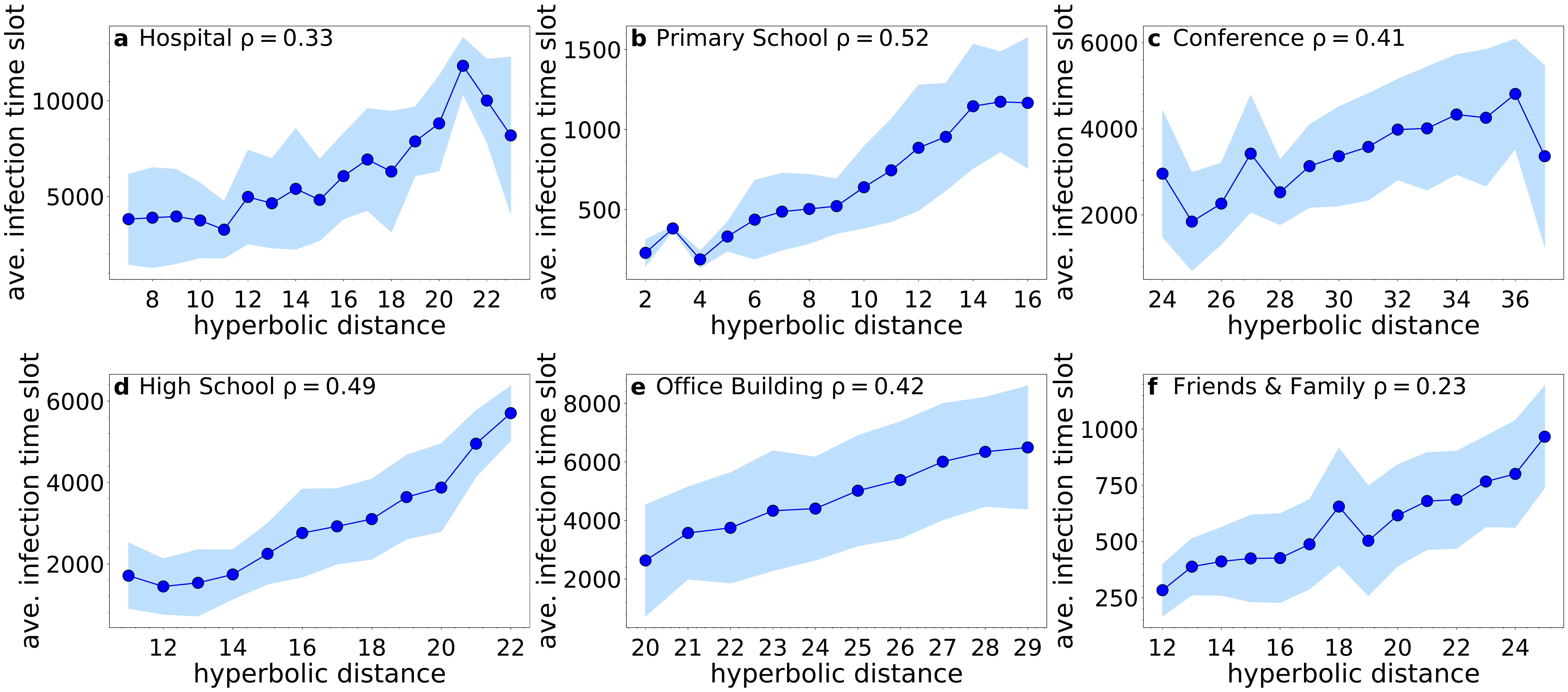}
\caption{Average infection time slot as a function of the hyperbolic distance from the infected source node in real networks. In each case we consider the inferred hyperbolic distances in the time-aggregated network formed over the full observation duration. The hyperbolic distance is binned into bins of size $\delta=1$ and the plots show the average infection time slot for nodes whose hyperbolic distance from the source node falls within each bin. The shaded area identifies the region corresponding to one standard deviation away from the average. Bins with less than $5$ samples are ignored. The results are averaged over $10$ simulated SI processes. Each process starts with a different infected source node selected at random, while the infection probability per time slot is $\beta=0.05$. Each plot indicates the average Spearman rank correlation coefficient $\rho$ between the infection time slot and the hyperbolic distance across the $10$ SI processes. In these plots we consider the hyperbolic distance instead of the equivalent effective distance $\tilde{\chi}$, as the former is more convenient for binning purposes.}
\label{fig:dp_real}
\end{figure}

\begin{figure}[t!]
\centering
\includegraphics[width=\linewidth]{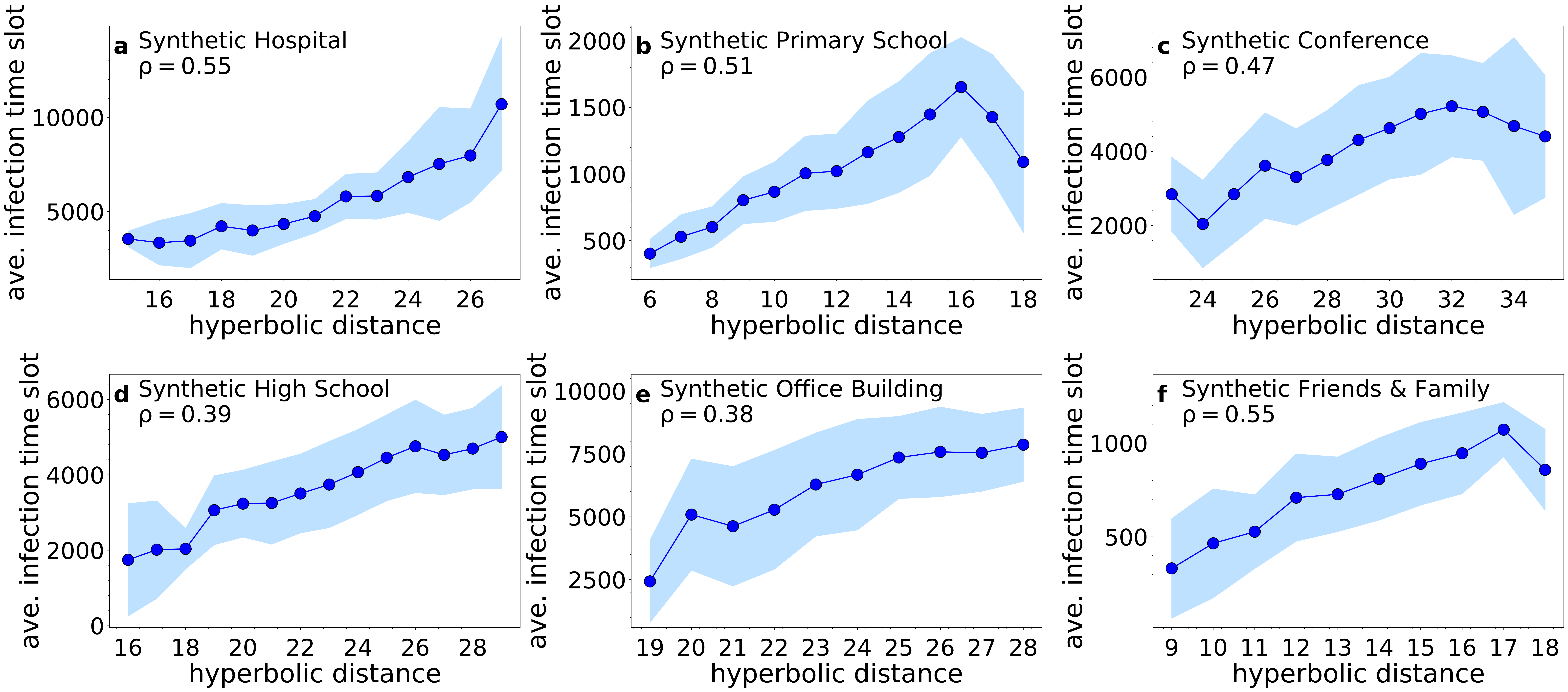}
\caption{Same as in Fig.~\ref{fig:dp_real} but for the synthetic counterparts (using inferred hyperbolic distances).}
\label{fig:dp_synt}
\end{figure}

Figs.~\ref{fig:dp_real} and~\ref{fig:dp_synt} show the results for the considered real networks and their synthetic counterparts, respectively. We see that the epidemic arrival times are significantly correlated with the hyperbolic distance from the infected source node. The correlation in each case is measured in terms of Spearman's rank correlation coefficient $\rho$ (see Appendix~\ref{sec:correlation}). These results indicate that hyperbolic embedding could provide a new perspective for understanding and predicting the behavior of epidemic spreading in human proximity systems. We leave further explorations for future work.

\section{Conclusion}

Individual snapshots of human proximity networks are often very sparse, consisting of a small number of interacting nodes. Nevertheless, we have shown that meaningful hyperbolic embeddings of such systems are still possible. Our approach is based on embedding the time-aggregated network of such systems over an adequately large observation period, using mapping methods developed for traditional complex networks. We have justified this approach by showing that the connection probability in the time-aggregated network is compatible with the Fermi-Dirac connection probability in random hyperbolic graphs, on which existing embedding methods are based. From an applications' perspective, we have shown that the hyperbolic maps of real proximity systems can be used to identify communities, facilitate efficient greedy routing on the temporal network, and predict future links. Further, we have shown that epidemic arrival times in the temporal network are positively correlated with the distance from the infection sources in the maps. Overall, our work opens the door for a geometric description of human proximity systems.

Our results indicate that the node coordinates change over time in the hyperbolic spaces of human proximity networks. An interesting yet challenging future work direction is to identify the stochastic differential equations that dictate this motion of nodes. Such equations would allow us to make predictions about the future positions of nodes in their hyperbolic spaces over different timescales. This, in turn, could allow us to improve the performance of tasks such as greedy routing and link prediction. This problem is relevant not only for human proximity systems, but for all complex networks where the hyperbolic node coordinates are expected to change over time, such as in social networks and the Internet~\cite{frag:hypermap_cn}. Another problem is to extend existing hyperbolic embedding methods so that they can refine the nodes' coordinates on a snapshot-by-snapshot basis as new snapshots become available, without having to recompute each time a new embedding from scratch. Such methods could be based on the idea that a local change in the system (new connections or disconnections) should involve mostly the neighborhood (coordinates of the nodes) around the change. For this purpose, techniques based on quadtree structures as in Ref.~\cite{henning2018} appear promising. Further, one might want to penalize large displacements based on the idea that the coordinates should be changing gradually from snapshot to snapshot. To this end,  Gaussian transition models for the coordinates as in Ref.~\cite{KimSurvey} seem appropriate. Methods for dynamic embedding in hyperbolic spaces should be useful not only for human proximity systems, but for temporal networks in general. 

\section*{Acknowledgements}

The authors acknowledge support by the TV-HGGs project (OPPORTUNITY/0916/ERC-CoG/0003), funded through the Cyprus Research and
Innovation Foundation.

\appendix

\section{Generating synthetic networks with the dynamic-$\mathbb{S}^1$ model}
\label{sec:ds1}

\begin{figure}[h!]
\centering
\includegraphics[width=0.8\linewidth]{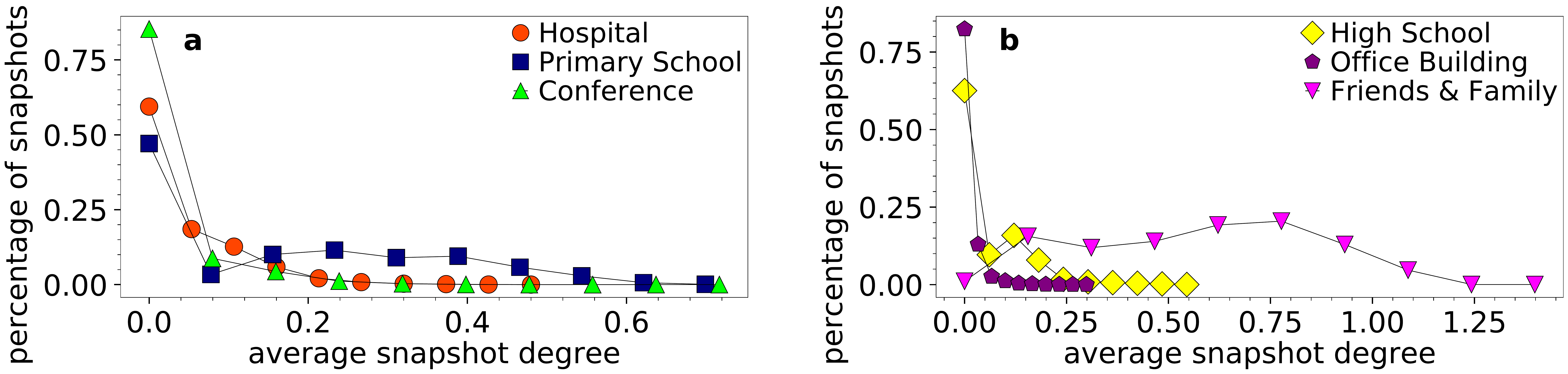}
\caption{Distribution of the average snapshot degree in the considered real networks.}
\label{fig:ave_kt}
\end{figure}

For each real network we construct its synthetic counterpart using the dynamic-$\mathbb{S}^{1}$ model as in Ref.~\cite{Papadopoulos2019}. Specifically, each counterpart has the same number of nodes $N$ and total duration (number of time slots) $\tau$ as the corresponding real network in Table~\ref{tab:networks}, while the latent variable $\kappa_i$ of each node $i=1, \ldots, N$ is set equal to the node's average degree per slot in the real network. The average degree $\bar{k}_t$ in each snapshot $G_t$, $t=1, \ldots, \tau$, is set equal to the average degree in the corresponding real snapshot at slot $t$---Fig.~\ref{fig:ave_kt} shows the distribution of $\bar{k}_t$. Finally, the temperature $T$ is set such that the resulting average time-aggregated degree, $\tilde{\bar{k}}$, is similar to the one in the real network. Each ``day'' in each counterpart corresponds to the same time slots as the corresponding day in the real system. See Ref.~\cite{Papadopoulos2019} for further details.

\section{Mercator}
\label{sec:mercator}

Mercator~\cite{GarciaPerez2019} combines the Laplacian Eigenmaps (LE) approach of Ref.~\cite{carlo1} with maximum likelihood estimation (MLE) to produce fast and accurate embeddings. It can embed networks with arbitrary degree distributions. In a nutshell, Mercator takes as input the network's adjacency matrix. It infers the nodes' latent degrees ($\tilde{\kappa}$) using the nodes' observed degrees in the network and the connection probability in the $\mathbb{S}^1$ model. To infer the nodes' angular coordinates ($\theta$), Mercator first utilizes the LE approach adjusted to the $\mathbb{S}^1$ model, in order to determine initial angular coordinates for the nodes. These initial angular coordinates are then refined using MLE, which adjusts the angular coordinates by maximizing the probability that the given network is produced by the  $\mathbb{S}^1$ model. Mercator also estimates the value of the temperature parameter $T$. The code implementing Mercator is made publicly available by the authors of~\cite{GarciaPerez2019} at~\url{https://github.com/networkgeometry/mercator}. We have used the code as is without any modifications.

As mentioned in the main text, we also considered a modified version of Mercator that replaces the connection probability of the $\mathbb{S}^1$ model in~(\ref{eq:fermi}) with the connection probability in~(\ref{eq:con_dS1_agg_v2}). This modification requires several changes to the original Mercator implementation that we describe in Appendix~\ref{sec:mod_mercator_app}.

\section{Epidemic arrival time and hyperbolic distance correlation}
\label{sec:correlation}

To quantify the correlation between the time slot at which a node becomes infected and its hyperbolic distance from the infected source node, we use Spearman's rank correlation coefficient $\rho$~\cite{Spearman1904}. Formally, given $n$ values $X_i$, $Y_i$, the values are converted to ranks $rg_{X_i}$, $rg_{Y_i}$, and Spearman's $\rho$ is computed as 
\begin{equation}
\rho=\frac{\mathrm{cov}(rg_X, rg_Y)}{\sigma_{rg_X}\sigma_{rg_Y}},
\end{equation}
where $\mathrm{cov}(rg_X, rg_Y)$ is the covariance of the rank variables, while $\sigma_{rg_X}, \sigma_{rg_Y}$ are the standard deviations of the rank variables. Spearman's $\rho$ takes values between $-1$ and $1$, and assesses monotonic relationships. $\rho=1$ ($\rho=-1$) occurs when there is a perfect monotonic increasing (decreasing) relationship between variables $X$ and $Y$, while $\rho=0$ indicates that there is no tendency for $Y$ to either increase or decrease when $X$ increases.


\section{Connection probability in the time-aggregated network}
\label{sec:connection_prob_app}

\begin{figure}[h!]
\centering
\includegraphics[width=\linewidth]{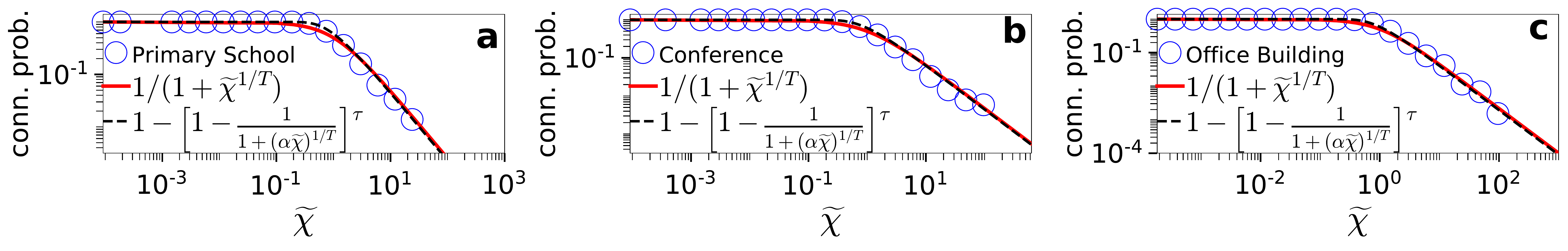}
\caption{\textbf{Connection probability in the time-aggregated network versus Fermi-Dirac connection probability.} Same as in Fig.~\ref{fig:fermi_vs_real} in the main text but for the synthetic counterparts of the primary school, conference and office building.}
\label{fig:fermi_vs_real_app}
\end{figure}

\section{Inference of latent coordinates with the original and modified Mercator}
\label{sec:inference_app}

\begin{figure}[h!]
\centering
\includegraphics[width=\linewidth]{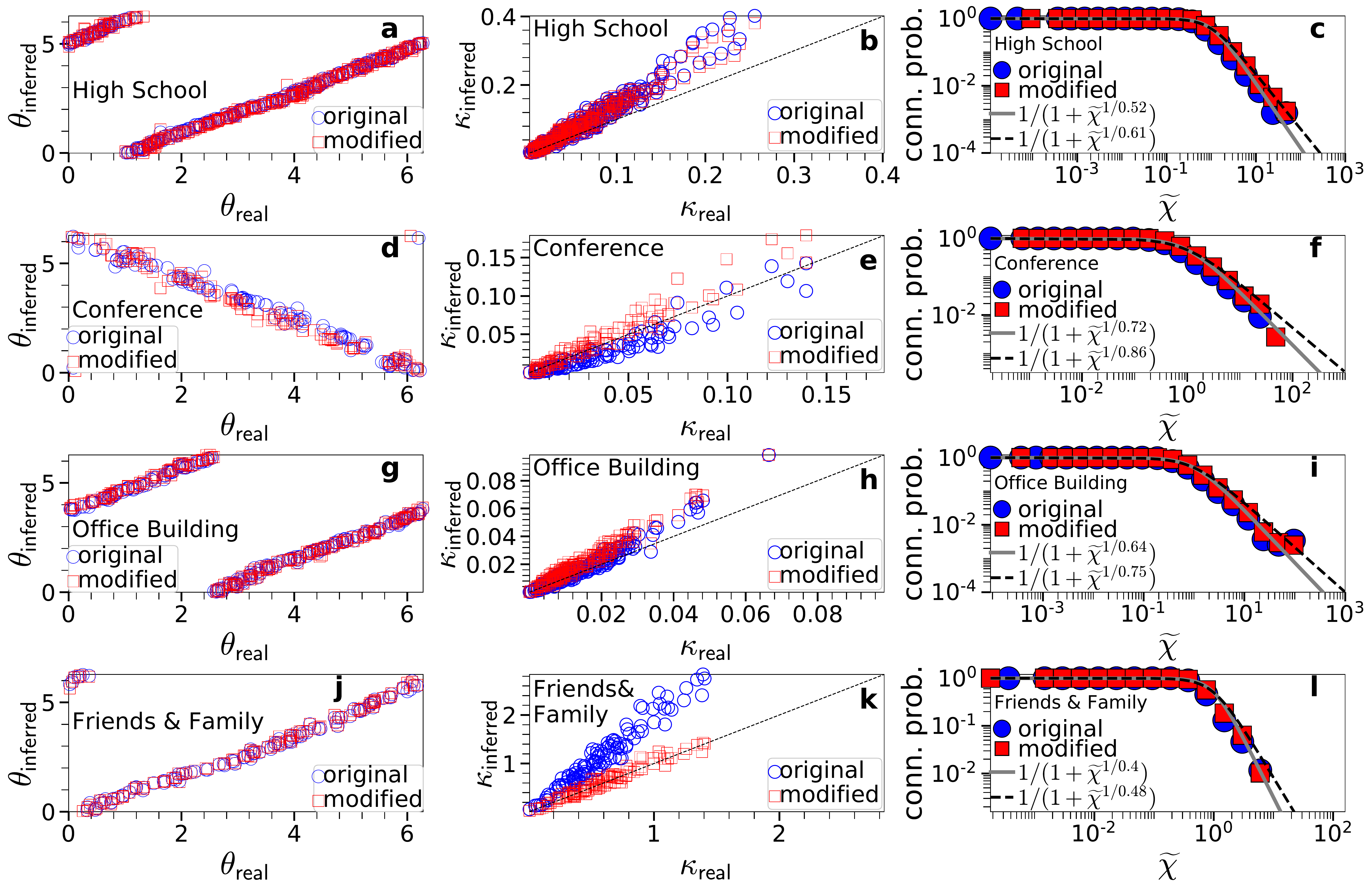}
\caption{\textbf{Inference of latent coordinates $(\kappa,\theta)$ with the original and modified versions of Mercator.} Same as in Fig.~\ref{fig:inf_mercator}  in the main text but for the synthetic counterparts of the high school, conference, office building and Friends \& Family. For the four networks, the original version estimates $T=0.52, 0.72, 0.64$ and $0.4$, the modified version estimates $T=0.61, 0.86, 0.75$ and $0.48$, while the actual values are $T=0.61, 0.85, 0.74$ and $0.48$, respectively.}
\label{fig:angles_k_si}
\end{figure}

\section{Hyperbolic maps of the conference and Friends \& Family}
\label{sec:maps_app}

\begin{figure}[h!]
\centering
\includegraphics[width=\linewidth]{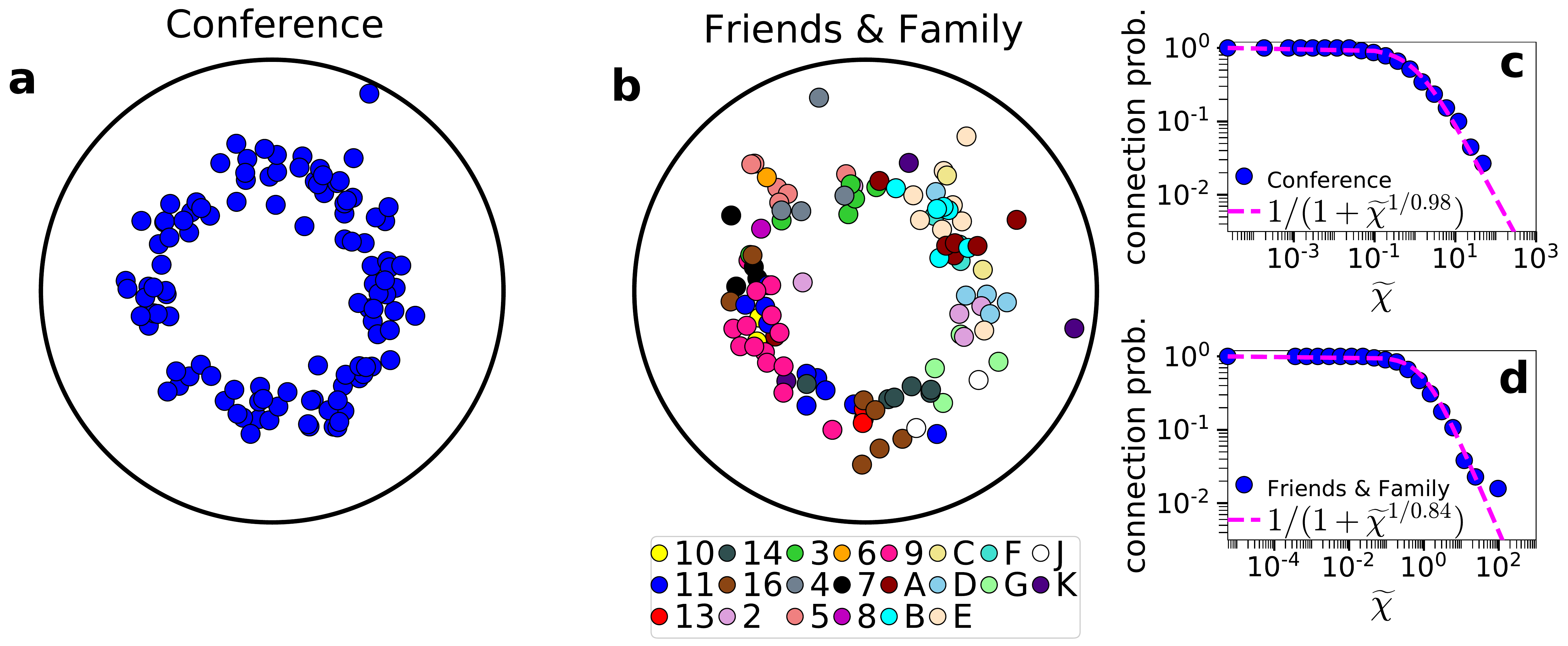}
\caption{\textbf{Hyperbolic embeddings of the conference and Friends \& Family.} Same as in Fig.~\ref{fig:hypermaps} in the main text but for the conference and Friends \& Family. In (\textbf{a}) all nodes have the same color as there is no group membership information available for the conference. In (\textbf{b}) the nodes are colored according to the partial apartment number or letter where they live, as given in the Friends \& Family metadata. The pink dashed lines in (\textbf{c}) and (\textbf{d}) are Fermi-Dirac connection probabilities with temperatures $T$ as inferred by Mercator, $T=0.98$ and $0.84$, respectively.}
\label{fig:hypermaps_si}
\end{figure}

\section{Human-to-human greedy routing}
\label{sec:h2h_gr_app}

\begin{figure}[h!]
\centering
\includegraphics[width=\linewidth]{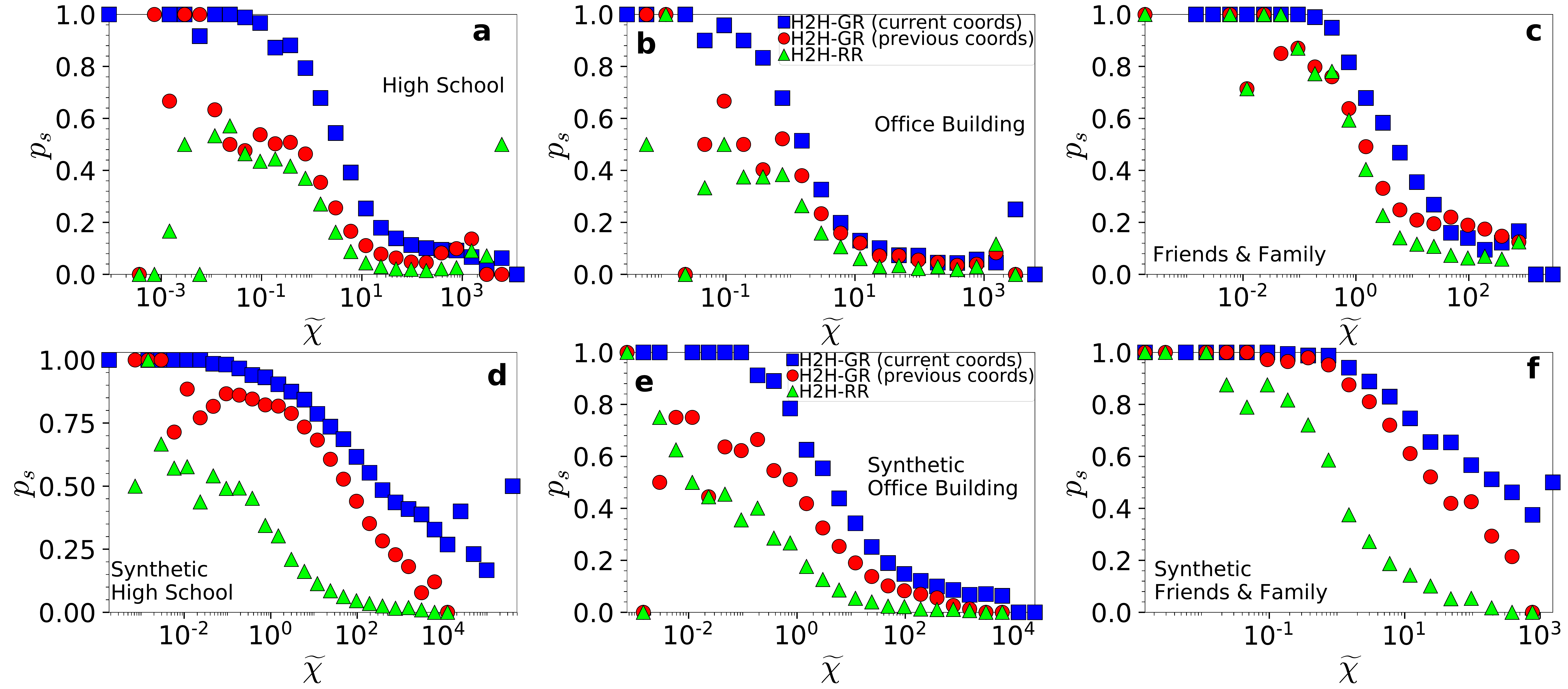}
\caption{\textbf{Success ratio $p_s$ of H2H-GR and H2H-RR as a function of the effective distance $\tilde{\chi}$ between source-destination pairs.} Same as in Fig.~\ref{fig:gr_success} in the main text but for the high school, office building and Friends \& Family. The top row shows the results for the real networks, while the bottom row shows the results for the synthetic counterparts.}
\label{fig:gr_si_success}
\end{figure}

\begin{figure}[h!]
\centering
\includegraphics[width=\linewidth]{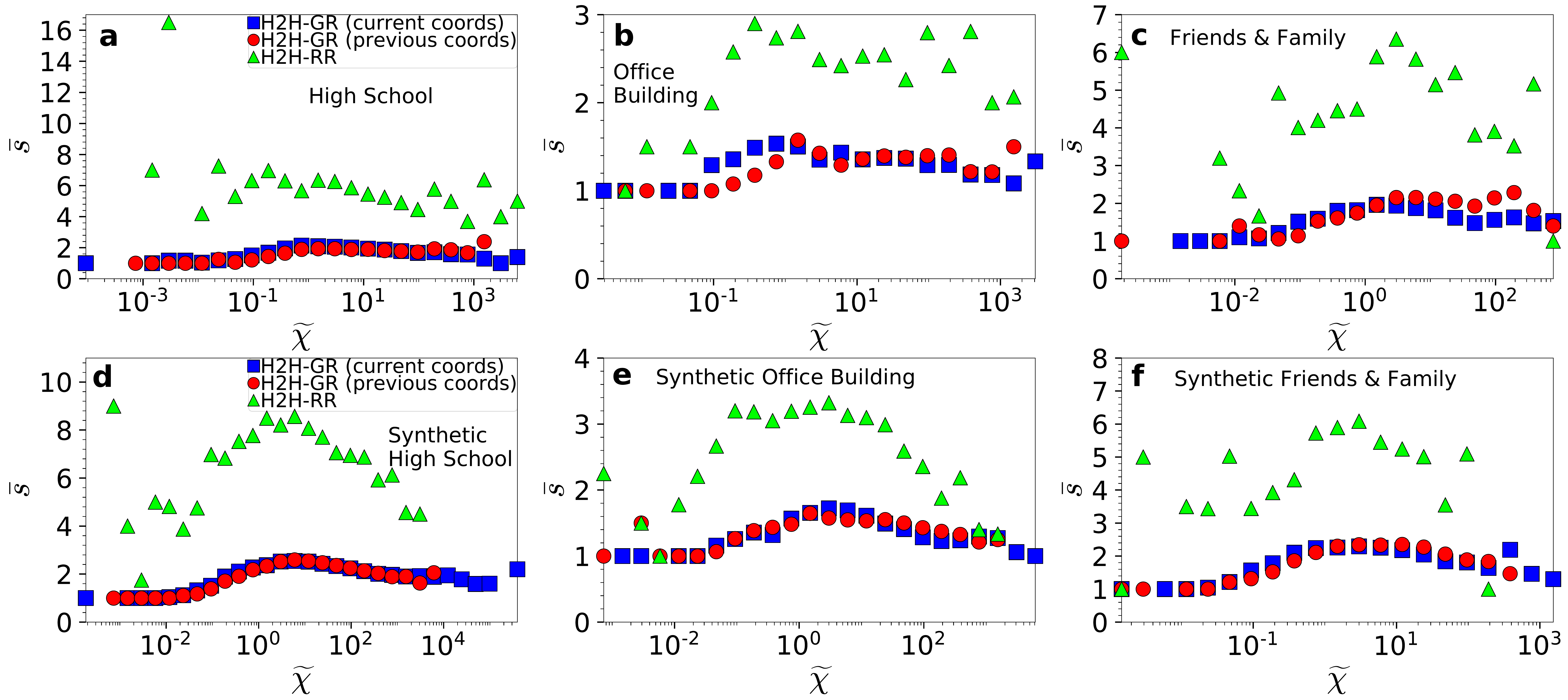}
\caption{\textbf{Average stretch $\bar{s}$ of H2H-GR and H2H-RR as a function of the effective distance $\tilde{\chi}$ between source-destination pairs.} The results correspond to the networks of Fig.~\ref{fig:gr_si_success}.}
\label{fig:gr_si_stretch}
\end{figure}

\begin{table}[ht!]
\centering
\begin{tabular}{|c|c|c|}
\hline
Real Network & H2H-GR (current angular coordinates) & H2H-GR (previous angular coordinates) \\
\hline
Hospital & $p_s=0.69$, $\bar{s}=2.16$ & $p_s=0.39$, $\bar{s}=1.98$ \\
\hline 
Primary School & $p_s=0.69$, $\bar{s}=4.40$ & $p_s=0.65$, $\bar{s}=3.88$ \\
\hline
Conference & $p_s=0.55$, $\bar{s}=2.25$ & $p_s=0.35$, $\bar{s}=2.11$ \\
\hline
High School & $p_s=0.20$, $\bar{s}=2.12$ & $p_s=0.10$, $\bar{s}=1.84$ \\
\hline
Office Building & $p_s=0.11$, $\bar{s}=1.42$ & $p_s=0.09$, $\bar{s}=1.39$ \\
\hline
Friends \& Family & $p_s=0.42$, $\bar{s}=2.20$ & $p_s=0.28$, $\bar{s}=2.03$ \\
\hline
\end{tabular}
\caption{\label{tab:gr_real_sim}\textbf{Success ratio $p_s$ and average stretch $\bar{s}$ of H2H-GR that uses only the angular (similarity) distances in real networks.} Same as in Table~\ref{tab:gr_real} in the main text but when using only the inferred angular coordinates (current and previous) in H2H-GR.}
\end{table}

\begin{table}[ht!]
\centering
\begin{tabular}{|c|c|c|}
\hline
Synthetic Network & H2H-GR (current angular coordinates) & H2H-GR (previous angular coordinates) \\
\hline
Hospital & $p_s=0.71$, $\bar{s}=2.46$ & $p_s=0.59$, $\bar{s}=2.44$ \\
\hline 
Primary School & $p_s=0.97$, $\bar{s}=4.77$ & $p_s=0.91$, $\bar{s}=5.29$ \\
\hline
Conference & $p_s=0.70$, $\bar{s}=2.83$ & $p_s=0.51$, $\bar{s}=2.77$ \\
\hline
High School & $p_s=0.35$, $\bar{s}=2.93$ & $p_s=0.25$, $\bar{s}=2.90$ \\
\hline
Office Building & $p_s=0.13$, $\bar{s}=1.66$ & $p_s=0.10$, $\bar{s}=1.61$ \\
\hline
Friends \& Family & $p_s=0.58$, $\bar{s}=2.58$ & $p_s=0.49$, $\bar{s}=2.47$ \\
\hline
\end{tabular}
\caption{\label{tab:gr_synth_sim}Same as in Table~\ref{tab:gr_real_sim} but for the synthetic counterparts of the real systems.}
\end{table}

\clearpage
\section{Stability of the inferred node coordinates in different days}
\label{sec:stability_app}

\begin{figure}[ht!]
\centering
\includegraphics[width=\linewidth, height=3.0in]{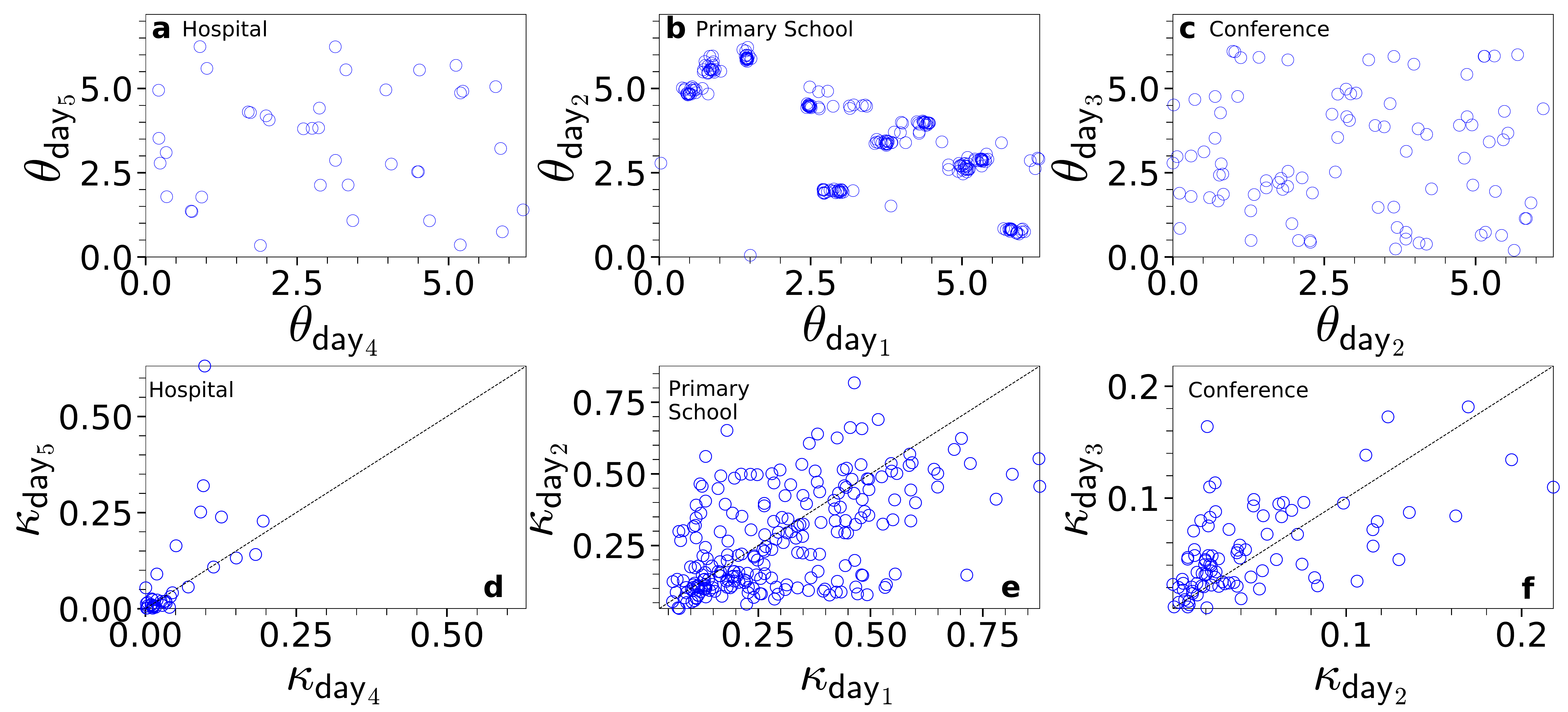}
\caption{\textbf{Inferred node coordinates $(\kappa, \theta)$ from the time-aggregated network of different observation days.} The results correspond to real networks, and the considered days are as in Table~\ref{tab:gr_real} in the main text. (\textbf{a}) Inferred angles in day~$4$ versus inferred angles in day~$5$ in the hospital. The numbers of time slots for days $4$ and $5$ are $\tau=3889$ and $2177$, respectively, while Mercator's inferred temperature for the time-aggregated network is $T=0.99$ for both days. (\textbf{b}) Inferred angles in day~$1$ versus inferred angles in day~$2$ in the primary school. For days $1, 2$, $\tau=1555, 1545$ and $T=0.43, 0.36$. (\textbf{c}) Inferred angles in day~$2$ versus inferred angles in day~$3$ in the conference. For days $2, 3$, $\tau=3216, 1946$ and $T=0.99, 0.98$. (\textbf{d}-\textbf{f}) Same as in (\textbf{a}-\textbf{c}) but for the inferred latent degrees $\kappa$. For each node, $\kappa$ is estimated as $\kappa=\tilde{\kappa}/\alpha$, where $\tilde{\kappa}$ is the node's inferred latent degree in the time-aggregated network of the corresponding day, while $\alpha=\tau^T/\Gamma(1+T)$. Due to rotational symmetry of the model, the inferred angles in a day can be globally shifted compared to the inferred angles in another day by any value in $[0, 2\pi]$.}
\label{fig:stability_real_a}
\end{figure}

\begin{figure}[hb!]
\centering
\includegraphics[width=\linewidth, height=3.0in]{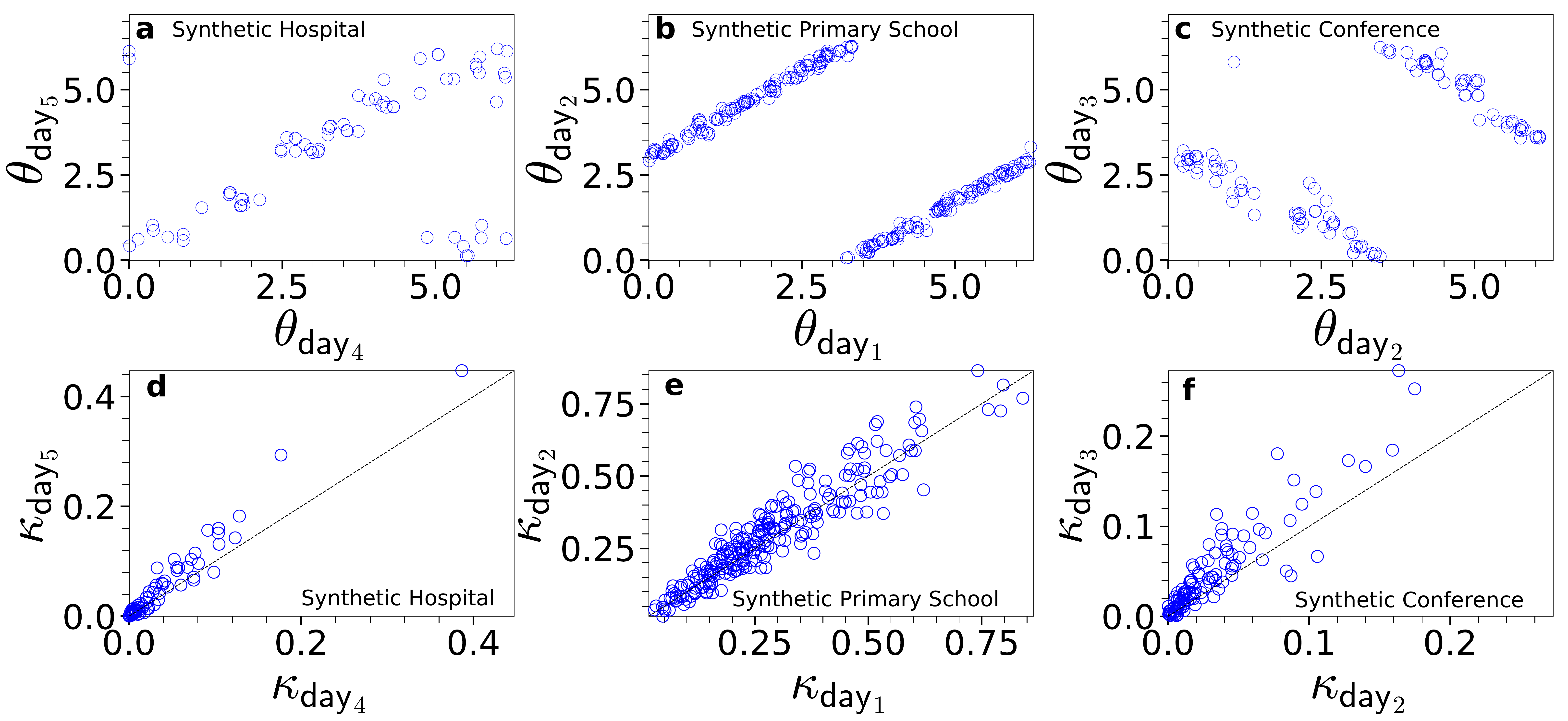}
\caption{\textbf{Inferred node coordinates $(\kappa, \theta)$ from the time-aggregated network of different days.} Same as in Fig.~\ref{fig:stability_real_a} but for the synthetic counterparts of the real systems. The days in each counterpart have the same duration $\tau$ as in the corresponding real system. The temperatures inferred by Mercator are $T=0.57$ for both days of the hospital, $T=0.60, 0.64$ for days $1, 2$ of the primary school, and $T=0.64$ for both days of the conference.}
\label{fig:stability_synth_a}
\end{figure}

\begin{figure}[ht!]
\centering
\includegraphics[width=\linewidth]{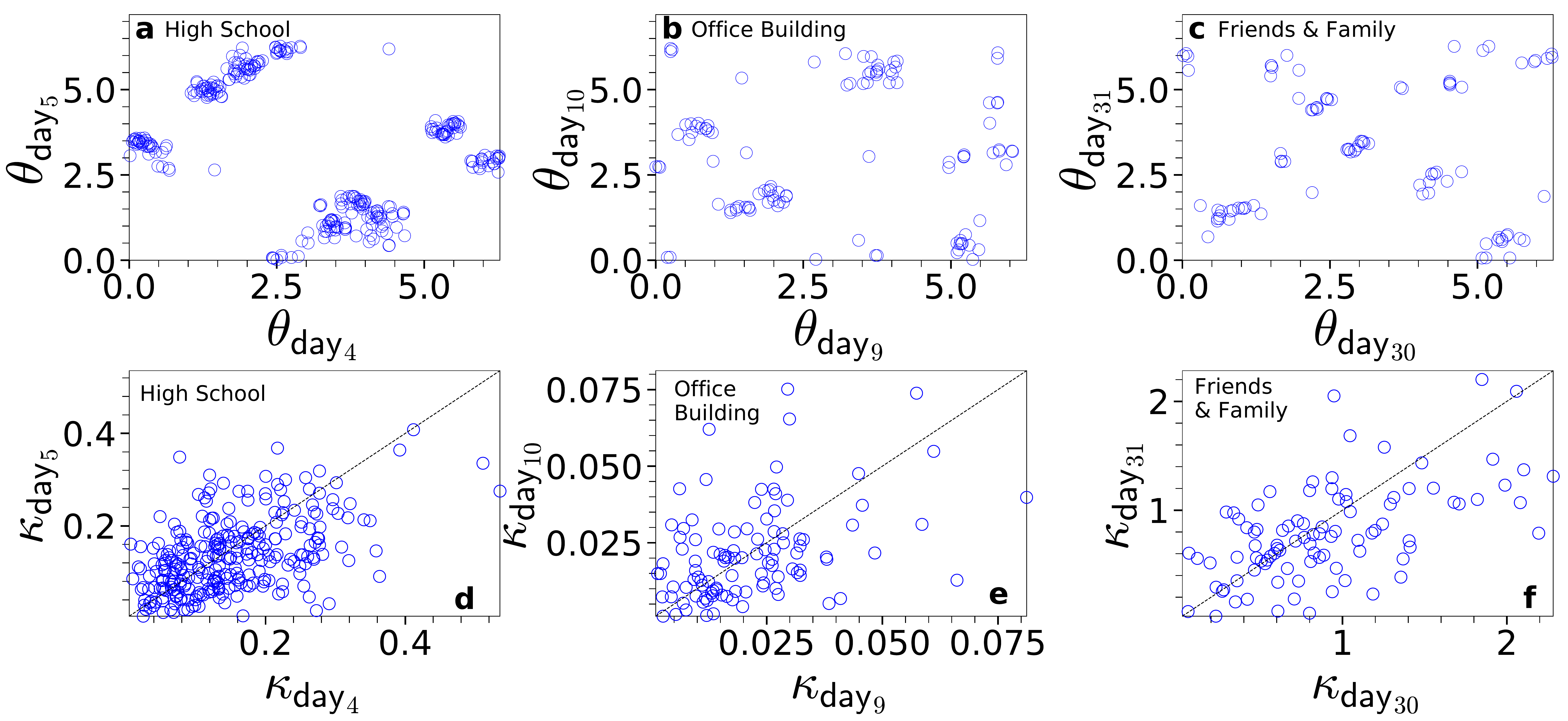}
\caption{\textbf{Inferred node coordinates $(\kappa, \theta)$ from the time-aggregated network of different observation days.} Same as in Fig.~\ref{fig:stability_real_a} but for the high school, office building and Friends \& Family. (\textbf{a}) Inferred angles in day~$4$ versus inferred angles in day~$5$ in the high school. For days $4, 5$, $\tau=1619$ and $T=0.54, 0.49$. (\textbf{b}) Inferred angles in day~$9$ versus inferred angles in day~$10$ in the office building. For days $9, 10$, $\tau=2153, 2148$ and $T=0.57, 0.47$. (\textbf{c}) Inferred angles in the $30^{\textnormal{th}}$ of March, 2011 versus inferred angles in the $31^{\textnormal{st}}$ of March, 2011 in the Friends \& Family. For the two days, $\tau=289$ and $T=0.52, 0.49$. (\textbf{d}-\textbf{f})~Same as in (\textbf{a}-\textbf{c}) but for the inferred latent degrees $\kappa$.}
\label{fig:stability_real_b}
\end{figure}

\begin{figure}[hb!]
\centering
\includegraphics[width=\linewidth]{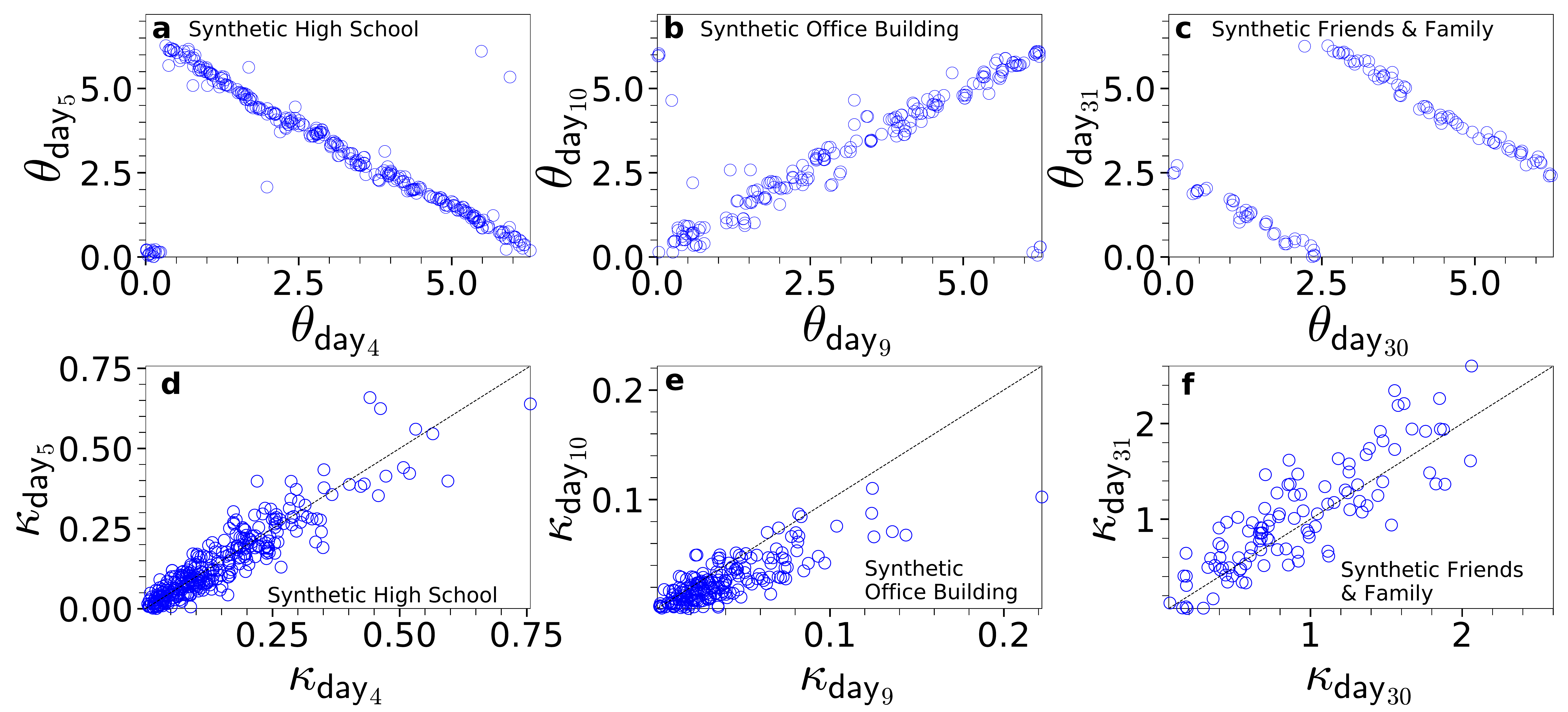}
\caption{\textbf{Inferred node coordinates $(\kappa, \theta)$ from the time-aggregated network of different days.} Same as in Fig.~\ref{fig:stability_real_b} but for the synthetic counterparts of the real systems. The days in each counterpart have the same duration $\tau$ as in the corresponding real system. The temperatures inferred by Mercator are $T=0.54$ for both days of the high school, $T=0.68$ for both days of the office building, and $T=0.45, 0.43$ for the two days of the Friends \& Family.} 
\label{fig:stability_synth_b}
\end{figure}

\begin{figure}[htb!]
\centering
\includegraphics[width=\linewidth]{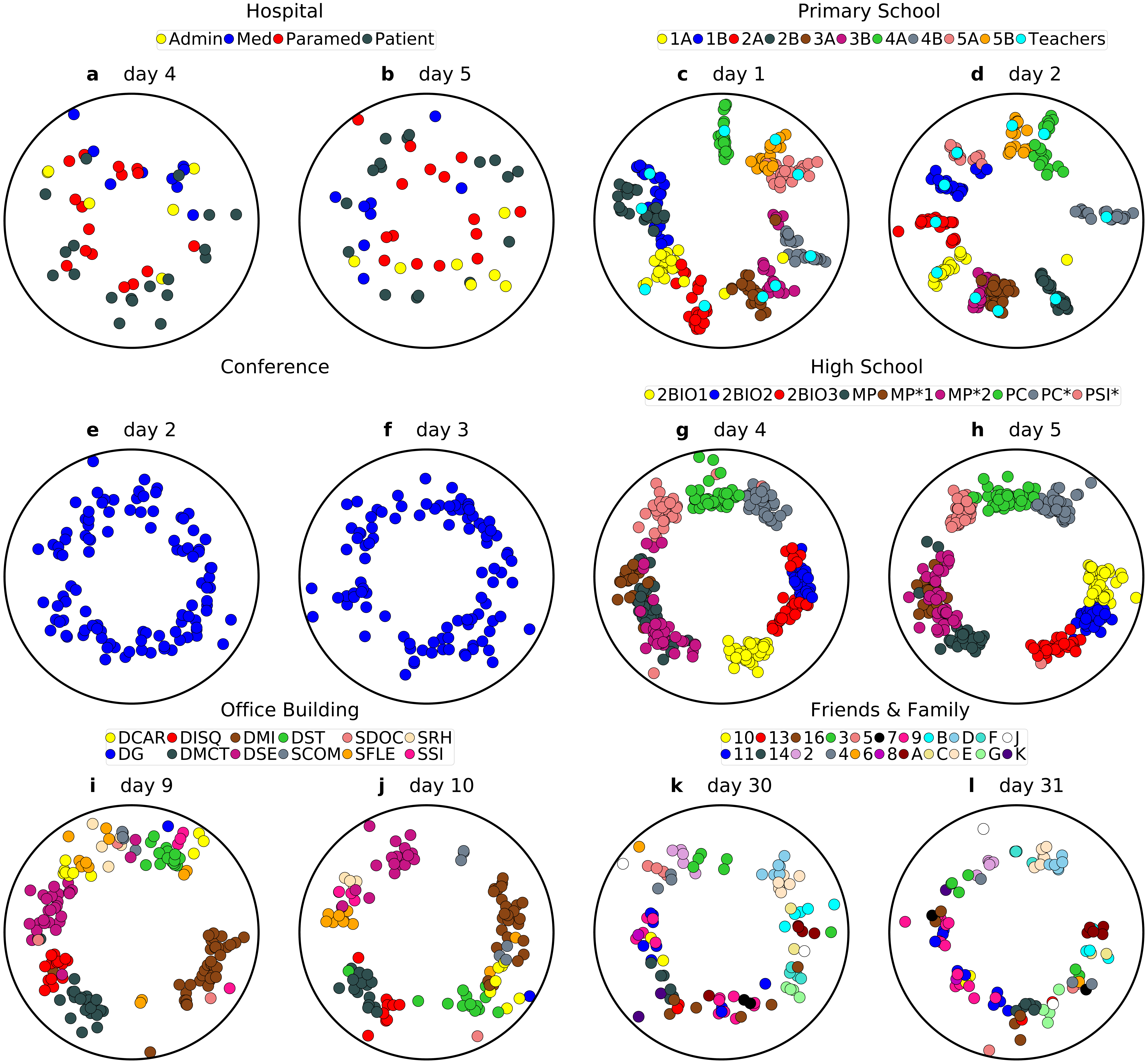}
\caption{\textbf{Daily hyperbolic maps of the considered real systems.} The maps correspond to the days considered in Figs.~\ref{fig:stability_real_a} and \ref{fig:stability_real_b}. The nodes are positioned according to their inferred hyperbolic coordinates $(r, \theta)$ in the time-aggregated network of the corresponding day. For an easier inspection of how node coordinates change between days, the map of each day is rotated such that it minimizes the sum of the squared distances between the inferred angles in the day and the angles inferred by considering the full duration of the corresponding network (Fig.~\ref{fig:hypermaps} in the main text and Fig.~\ref{fig:hypermaps_si}). The nodes are colored according to group membership information available in the metadata of each network as described in Fig.~\ref{fig:hypermaps} of the main text and in Fig.~\ref{fig:hypermaps_si}.
\label{fig:stability_maps}}
\end{figure}

\section{Modified Mercator}
\label{sec:mod_mercator_app}

In the modified version of Mercator we replace the connection probability of the $\mathbb{S}^1$ model [Eq.~(\ref{eq:fermi}) in the main text] with the connection probability in the time-aggregated network of the dynamic-$\mathbb{S}^1$ model [Eq.~(\ref{eq:con_dS1_agg_v2}) in the main text]. This modification requires replacing all relations in the original Mercator implementation, which are derived using the original connection probability, with the corresponding relations derived using the new connection probability. Below we list all modifications made in each step of the original Mercator implementation (Ref.~\cite{GarciaPerez2019}). For convenience, we express all new relations in terms of the nodes' latent degrees per snapshot $\kappa$. We recall that $\kappa=\tilde{\kappa}/\alpha$, where $\tilde{\kappa}$ denotes the node's latent degree in the time-aggregated network, while $\alpha=\tau^T/\Gamma(1+T)$. The modified Mercator takes also as input the value of the duration $\tau$ in which the time-aggregated network is computed, while like the original version it infers the value of the temperature parameter $T$ along with the nodes' coordinates ($\tilde{\kappa}, \theta$).

In the first step, Mercator uses an iterative procedure that adjusts the nodes' latent degrees so that the expected degree of each node as prescribed by the $\mathbb{S}^1$ model matches the node's observed degree in the given network. We adapt this step by replacing the relation for the probability that two nodes with latent degrees $\kappa$ and $\kappa'$ are connected [Eq.~(A1) in Ref.~\cite{GarciaPerez2019}], with
\begin{align}
\label{eq:a1}
p(a_{\kappa\kappa'}=1) = 1- \frac{2\mu\kappa\kappa'}{N}\left[\frac{T\Gamma(\tau+T)\Gamma(-T)}{\Gamma(\tau)} + \left(\frac{1}{1+\left(\frac{N}{2\mu\kappa\kappa'} \right)^{1/T}} \right)^{-T} {}_2F_1\left(-T,1-\tau-T; 1-T; \frac{1}{1+\left(\frac{N}{2\mu\kappa\kappa'} \right)^{1/T}}\right) \right].
\end{align}
The above relation is derived in Ref.~\cite{Papadopoulos2019}. $a_{\kappa\kappa'}$ is an indicator function, $a_{\kappa\kappa'}=1$ if two nodes with latent degrees $\kappa$ and $\kappa'$ are connected in the time-aggregated network, and $a_{\kappa\kappa'}=0$ otherwise, $\mu=\sin(T\pi)/(2\bar{\kappa}T\pi)$, and ${}_2F_1(a, b; c; z)$ is the Gauss hypergeometric function.

In the second step, Mercator uses an iterative procedure that adjusts the temperature $T$ so that the value of the average clustering coefficient as prescribed by the $\mathbb{S}^1$ model matches the value of the average clustering coefficient in the given network. We adapt this step by replacing the relation for the distribution of the angular distance $\Delta \theta$ between two connected nodes with latent degrees $\kappa$ and $\kappa'$ [Eq.~(A3) in Ref.~\cite{GarciaPerez2019}], with
\begin{align}
\label{eq:a3}
\nonumber \rho(\Delta\theta\vert a_{\kappa\kappa'}=1)&=\frac{p(a_{\kappa\kappa'}=1\vert\Delta\theta)\rho(\Delta\theta)}{p(a_{\kappa\kappa'}=1)} \\
&=\frac{\frac{1}{\pi}\left[1-\left(1 - \frac{1}{1 + \left(\frac{R\Delta\theta}{\mu\kappa\kappa'}\right)^{1/T}}\right)^\tau\right]}{1- \frac{2\mu \kappa\kappa'}{N}\left[\frac{T\Gamma(\tau+T)\Gamma(-T)}{\Gamma(\tau)} + \left(\frac{1}{1+\left(\frac{N}{2\mu\kappa\kappa'} \right)^{1/T}} \right)^{-T} {}_2F_1\left(-T,1-\tau-T; 1-T; \frac{1}{1+\left(\frac{N}{2\mu\kappa\kappa'} \right)^{1/T}}\right) \right]}.
\end{align} 
In the above relation, $p(a_{\kappa\kappa'}=1\vert\Delta\theta)$ is the probability that two nodes with latent degrees $\kappa$ and $\kappa'$ and angular distance $\Delta\theta$ are connected in the time-aggregated network, $\rho(\Delta\theta)=1/\pi$ is the uniform distribution of the angular distances in the model, and $p(a_{\kappa\kappa'}=1)$ is given by~(\ref{eq:a1}). 

In the third step, Mercator adapts Laplacian Eigenmaps (LE) to the $\mathbb{S}^1$ model in order to determine initial angular coordinates for the nodes. We adapt this step by replacing the relation for the expected angular distance between two nodes with latent degrees $\kappa_i$ and $\kappa_j$ conditioned on the fact that they are connected [Eq.~(A8) in Ref.~\cite{GarciaPerez2019}], with
\begin{align}
\label{eq:a8}
\nonumber\left\langle \Delta\theta_{ij} \right\rangle &= \int_0^\pi \Delta\theta_{ij}\rho(\Delta\theta_{ij}\vert a_{\kappa_i\kappa_j}=1)\mathrm{d}\Delta\theta_{ij} \\
&= \frac{N\pi\Gamma(\tau )\left[\tau+2T-2T\left(\frac{\pi R}{\mu \kappa_i\kappa_j}\right)^{\tau /T} \, _2F_1\left(\tau , \tau+2T; \tau+2T+1; -\left(\frac{\pi R}{\mu \kappa_i\kappa_j}\right)^{1/T}\right)\right]}{2 (\tau+2T) \Bigg[\Bigg(N+2T\mu \kappa_i\kappa_j \textnormal{B}(x; -T, \tau+T)\Bigg)\Gamma(\tau)+2\mu \kappa_i\kappa_j \Gamma(1-T) \Gamma(\tau+T)\Bigg]},
\end{align}
where $\textnormal{B}(x; a, b)$ is the incomplete beta function and $x=1/\left[1+\left(\frac{N}{2\mu\kappa_i\kappa_j} \right)^{1/T}\right]$. In this step, we also replace the probability in the $\mathbb{S}^1$ model of having the observed connection (or disconnection) among each pair of consecutive nodes $i$ and $i+1$ on the similarity circle, conditioned on their angular separation gap $g_i$ and their latent degrees $\kappa_i$ and $\kappa_j$ [Eq.~(A13) in Ref.~\cite{GarciaPerez2019}], with
\begin{align}
\label{eq:a13}
p(a_{i+1,i}\vert g_i)=\left[1-\left(1 - \frac{1}{1 + \left(\frac{R g_i}{\mu\kappa_i\kappa_j}\right)^{1/T}}\right)^\tau \right]^{a_{i+1,i}}\times \left[\left(1 - \frac{1}{1 + \left(\frac{R g_i}{\mu\kappa_i\kappa_j}\right)^{1/T}}\right)^\tau \right]^{1-a_{i+1,i}},
\end{align}
where $a_{i, i+1}=1$ if the two nodes are connected in the time-aggregated network, and $a_{i, i+1}=0$ otherwise.

In the fourth step, Mercator refines the initial angular coordinates by (approximately) maximizing the likelihood that the given network is produced by the $\mathbb{S}^1$ model. We adapt this step by replacing the local log-likelihood for each node $i$ [Eq.~(A20) in Ref.~\cite{GarciaPerez2019}], with
\begin{equation}
\ln\mathcal{L}_i = \sum_{j\neq i} a_{ij}\ln\left[1-\left(1 - \frac{1}{1 + \left(\frac{R\Delta\theta_{ij}}{\mu\kappa_i\kappa_j}\right)^{1/T}}\right)^\tau\right] + (1-a_{ij})\ln\left[\left(1 - \frac{1}{1 + \left(\frac{R\Delta\theta_{ij}}{\mu\kappa_i\kappa_j}\right)^{1/T}}\right)^\tau\right],
\end{equation}
where $a_{i j}=1$ if nodes $i$ and $j$ are connected in the time-aggregated network, and $a_{i j}=0$ otherwise.

In the final (optional) step, Mercator re-adjusts the latent degrees of the nodes according to the inferred angular coordinates so that the expected degree of each node indeed matches its observed degree in the given network. We adapt this step by replacing the connection probability of the $\mathbb{S}^1$ model in Eq.~(A21) in Ref.~\cite{GarciaPerez2019}, with the connection probability in the time-aggregated network of the dynamic-$\mathbb{S}^1$ model [Eq.~(\ref{eq:con_dS1_agg_v2}) in the main text].

\section{Aggregation interval and rotation of angular coordinates}
\label{sec:aggr_and_rot}

\begin{figure}[htb!]
\vspace{-0.06in}
\centering
\includegraphics[width=\linewidth]{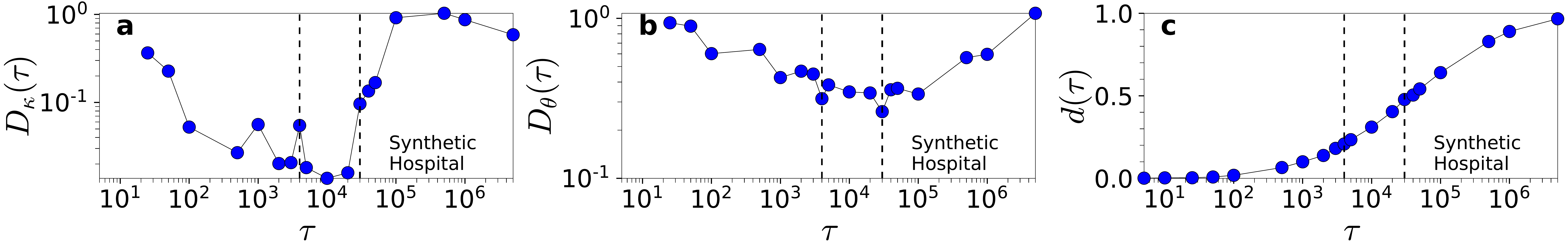}
\caption{\textbf{Inference accuracy~vs.~aggregation interval.} Same as in Fig.~\ref{fig:aggregation_effect} in the main text but for a synthetic counterpart of the hospital. The vertical dashed lines indicate the interval $4000 \leq \tau \leq 30000$. In this interval, $D_{\kappa}(\tau) < 0.1$, $D_{\theta}(\tau) < 0.4$, and $0.20 < d(\tau) < 0.48$.
\label{fig:interval_hp}}
\end{figure}

\begin{figure}[htb!]
\vspace{-0.06in}
\centering
\includegraphics[width=\linewidth]{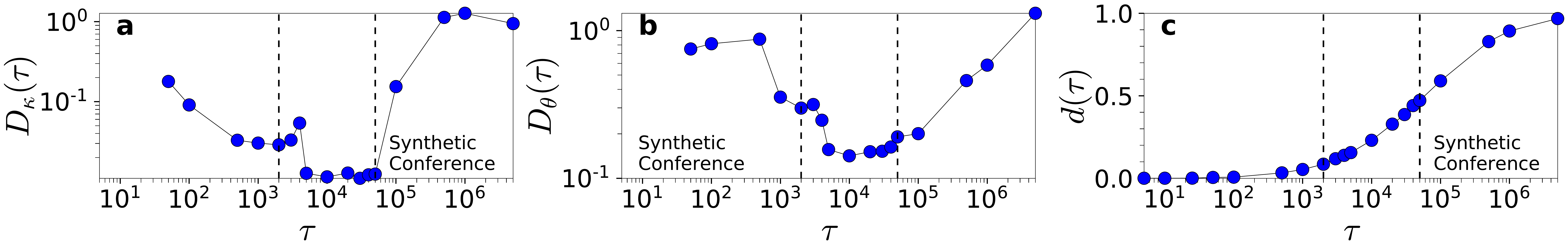}
\caption{\textbf{Inference accuracy~vs.~aggregation interval.} Same as in Fig.~\ref{fig:interval_hp} but for a synthetic counterpart of the conference. The vertical dashed lines indicate the interval $2000 \leq \tau \leq 50000$. In this interval, $D_{\kappa}(\tau)< 0.1$, $D_{\theta}(\tau)< 0.4$, and $0.09 < d(\tau) < 0.47$.
\label{fig:interval_cf}}
\end{figure}

\begin{figure}[htb!]
\vspace{-0.06in}
\centering
\includegraphics[width=\linewidth]{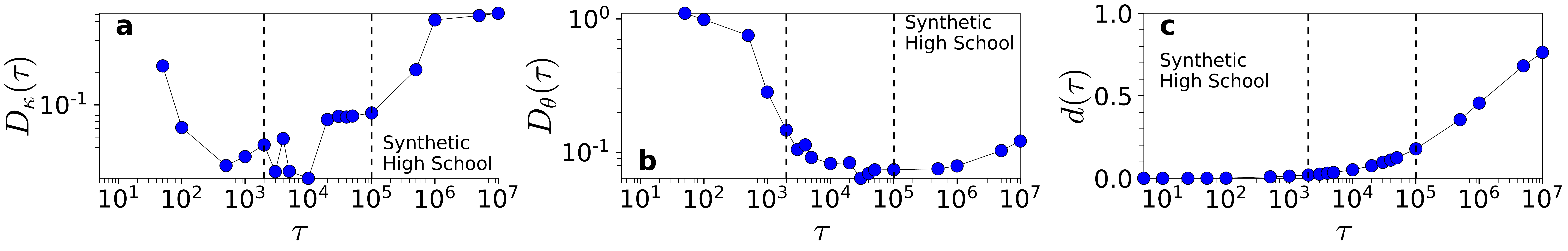}
\caption{\textbf{Inference accuracy~vs.~aggregation interval.} Same as in Fig.~\ref{fig:interval_hp} but for a synthetic counterpart of the high school. The vertical dashed lines indicate the interval $2000 \leq \tau \leq 100000$. In this interval, $D_{\kappa}(\tau) < 0.1$, $D_{\theta}(\tau) < 0.2$, and $0.01 < d(\tau) < 0.18$.
\label{fig:interval_hs}}
\end{figure}

\begin{figure}[htb!]
\vspace{-0.06in}
\centering
\includegraphics[width=\linewidth]{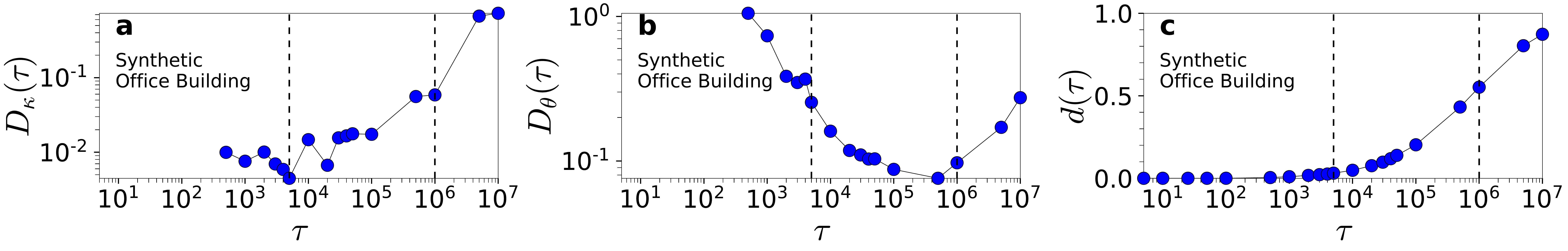}
\caption{\textbf{Inference accuracy~vs.~aggregation interval.} Same as in Fig.~\ref{fig:interval_hp} but for a synthetic counterpart of the office building. The vertical dashed lines indicate the interval $5000 \leq \tau \leq 1000000$. In this interval, $D_{\kappa}(\tau) < 0.1$, $D_{\theta}(\tau) < 0.3$, and $0.03 < d(\tau) < 0.55$.
\label{fig:interval_of}}
\end{figure}

\begin{figure}[htb!]
\vspace{-0.06in}
\centering
\includegraphics[width=\linewidth]{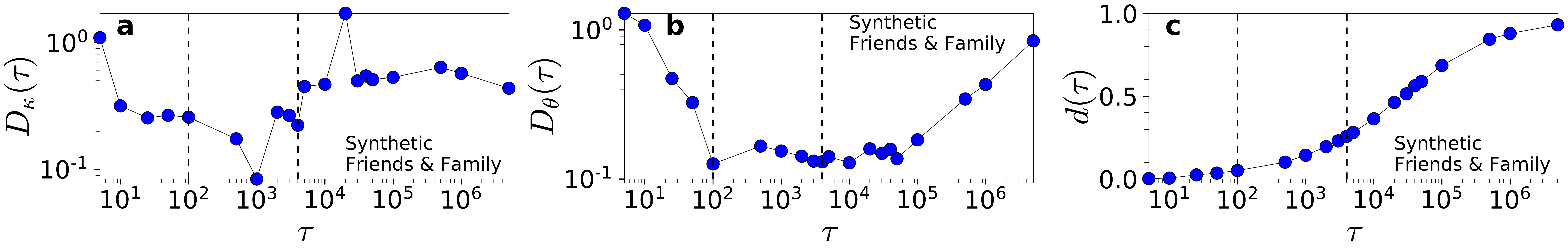}
\caption{\textbf{Inference accuracy~vs.~aggregation interval.} Same as in Fig.~\ref{fig:interval_hp} but for a synthetic counterpart of the friends \& family. The vertical dashed lines indicate the interval $100 \leq \tau \leq 4000$. In this interval, $D_{\kappa}(\tau)<0.3$, $D_{\theta}(\tau) < 0.2$, and $0.05 < d(\tau) < 0.26$.
\label{fig:interval_ff}}
\end{figure}

\begin{figure}[htb!]
\centering
\includegraphics[width=\linewidth]{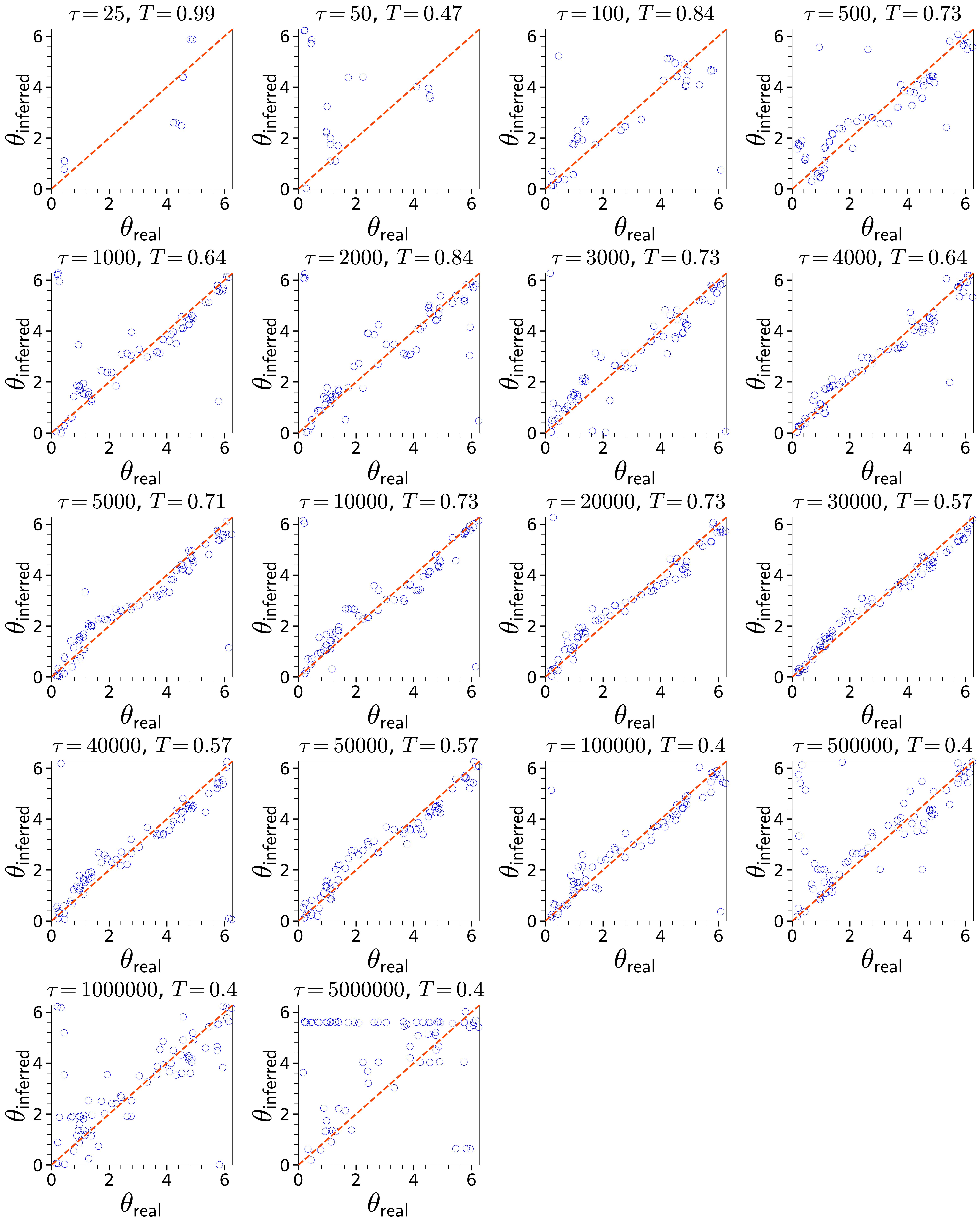}
\caption{\textbf{Inferred~vs.~real $\theta$ for different aggregation intervals $\tau$.} The results correspond to the synthetic counterpart of the hospital. For each $\tau$ we also indicate the temperature $T$ inferred by Mercator. The diagonal dashed line indicates $x=y$.
\label{fig:thetas_diff_intervals_hp}}
\end{figure}

\begin{figure}[htb!]
\centering
\includegraphics[width=\linewidth]{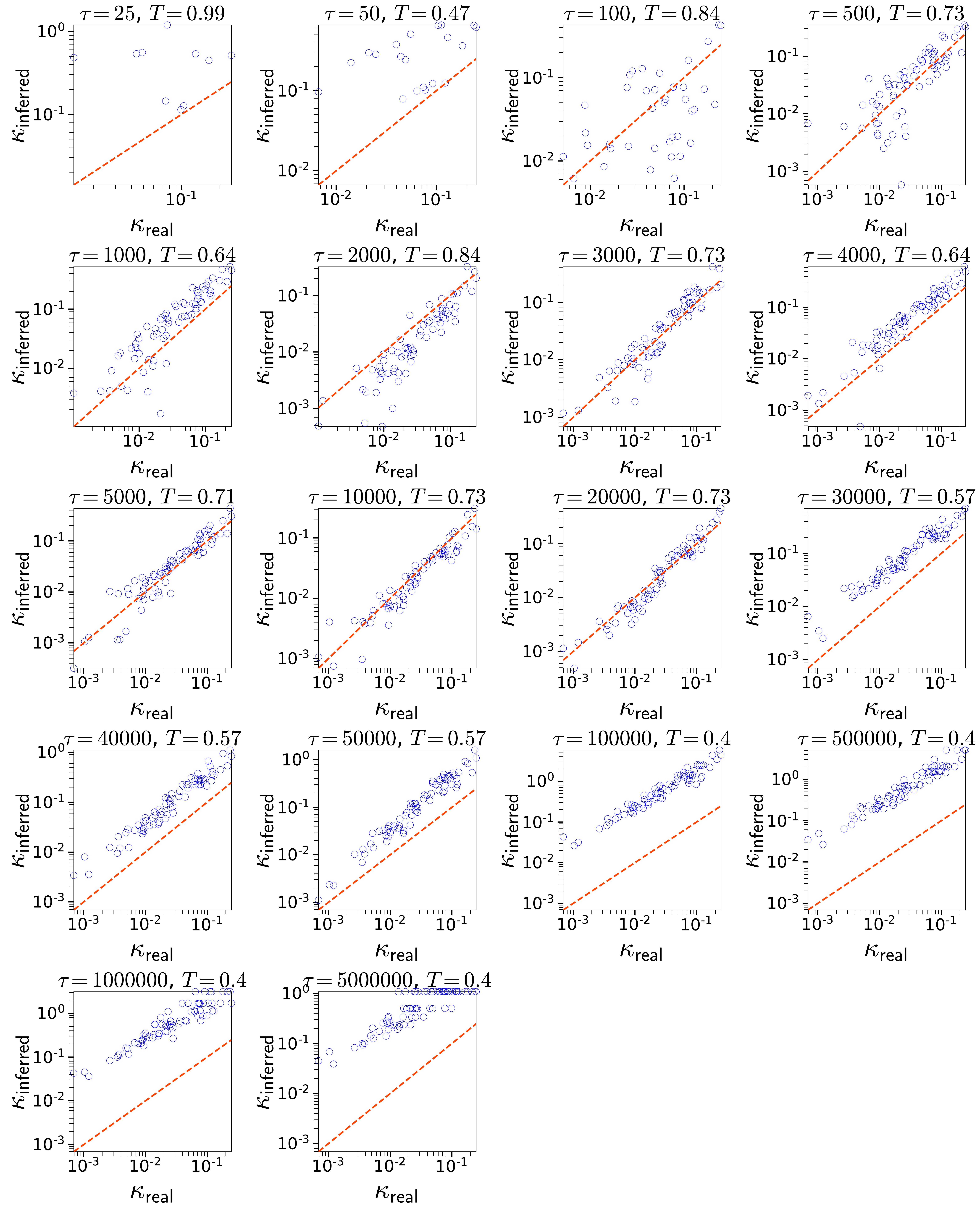}
\caption{\textbf{Inferred~vs.~real $\kappa$ for different aggregation intervals $\tau$.} The results correspond to the synthetic counterpart of the hospital. The $\kappa_{\textnormal{inferred}}$ are estimated as described in the caption of Fig.~\ref{fig:inf_mercator} in the main text. For each $\tau$ we indicate the temperature $T$ inferred by Mercator. The diagonal dashed line indicates $x=y$.
\label{fig:kappas_diff_intervals_hp}}
\end{figure}

\begin{figure}[htb!]
\centering
\includegraphics[width=\linewidth]{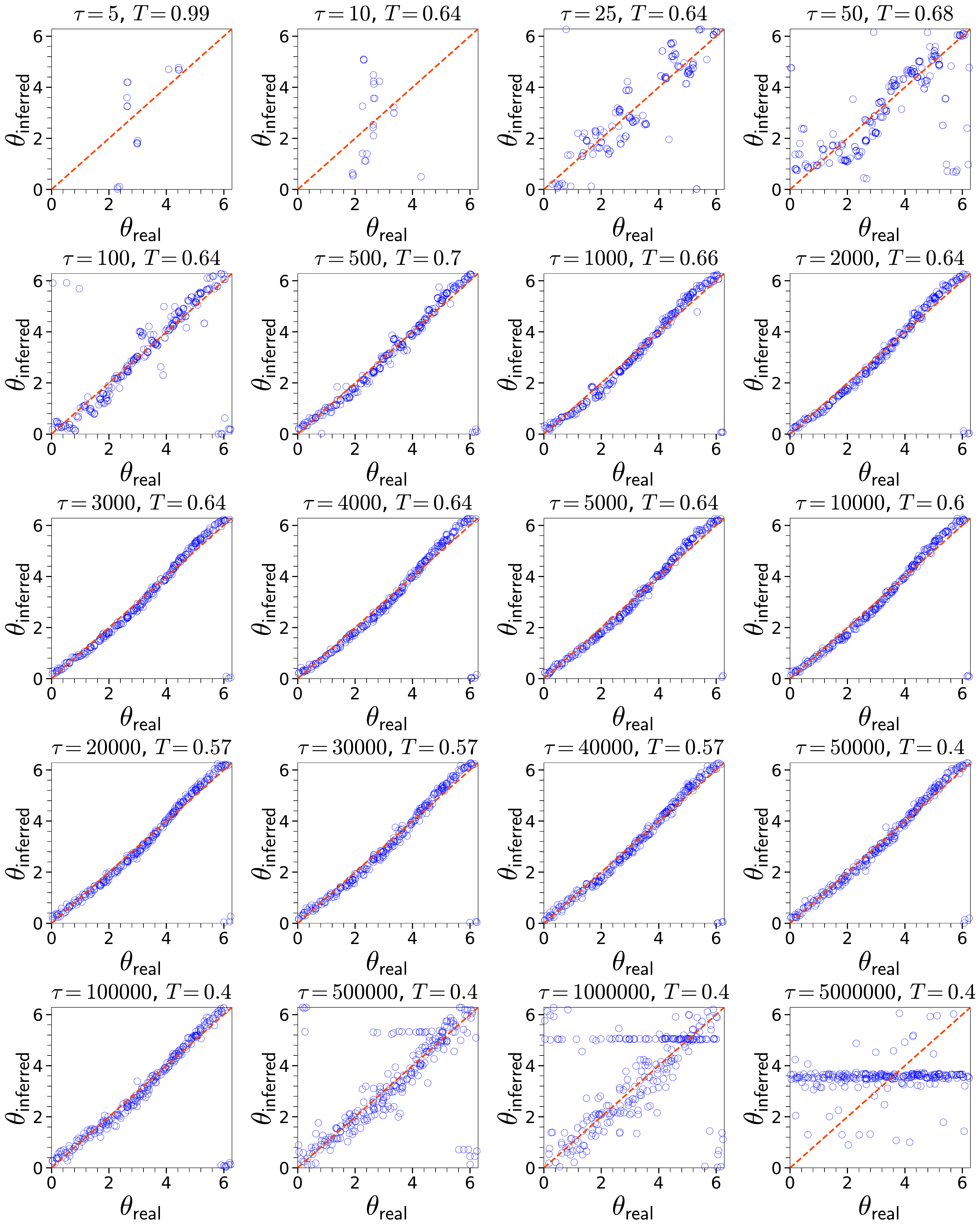}
\caption{\textbf{Inferred~vs.~real $\theta$ for different aggregation intervals $\tau$.} Same as in Fig.~\ref{fig:thetas_diff_intervals_hp} but for the synthetic counterpart of the primary school.
\label{fig:thetas_diff_intervals_ps}}
\end{figure}

\begin{figure}[htb!]
\centering
\includegraphics[width=\linewidth]{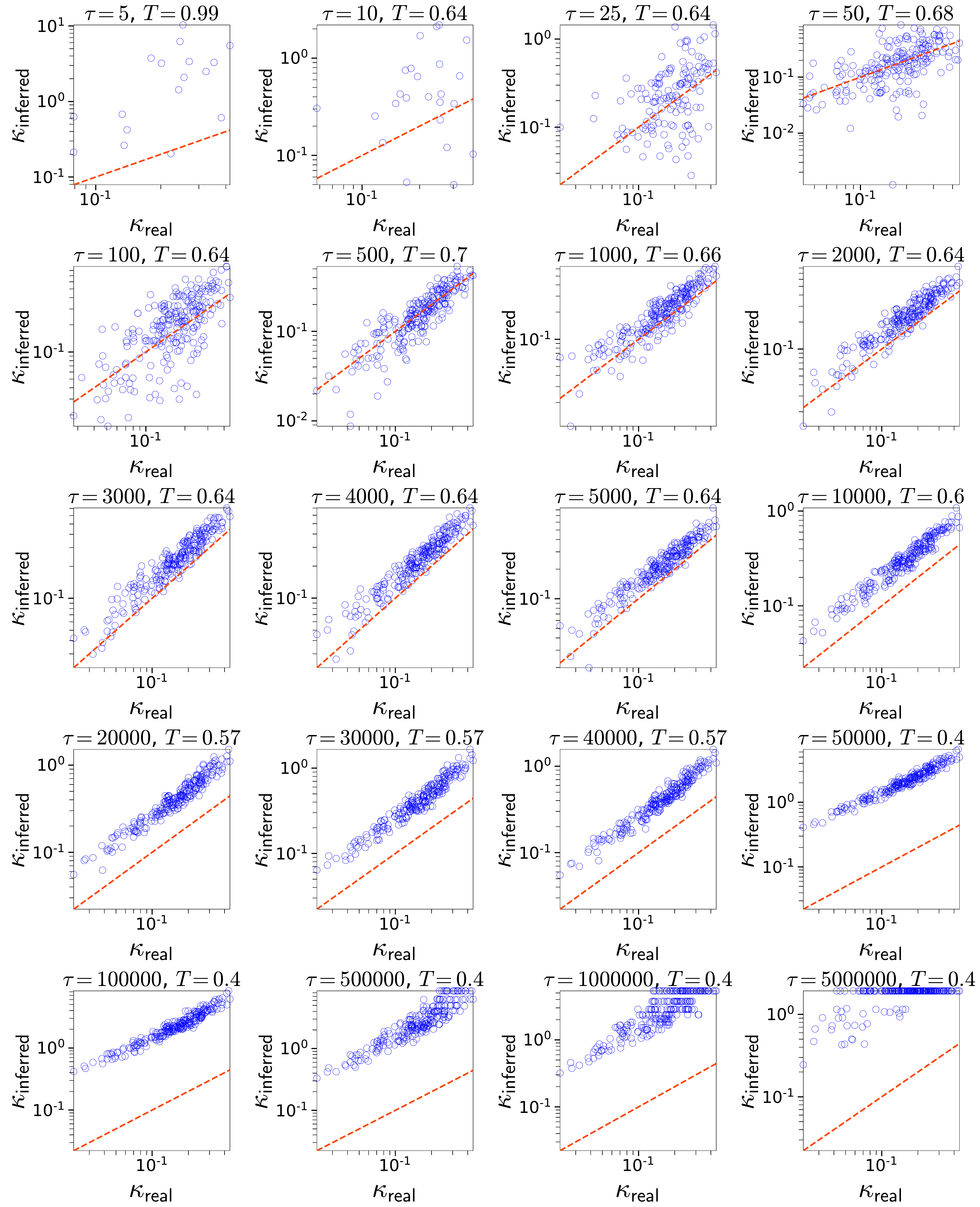}
\caption{\textbf{Inferred~vs.~real $\kappa$ for different aggregation intervals $\tau$.} Same as in Fig.~\ref{fig:kappas_diff_intervals_hp} but for the synthetic counterpart of the primary school.
\label{fig:kappas_diff_intervals_ps}}
\end{figure}

\begin{figure}[htb!]
\centering
\includegraphics[width=\linewidth]{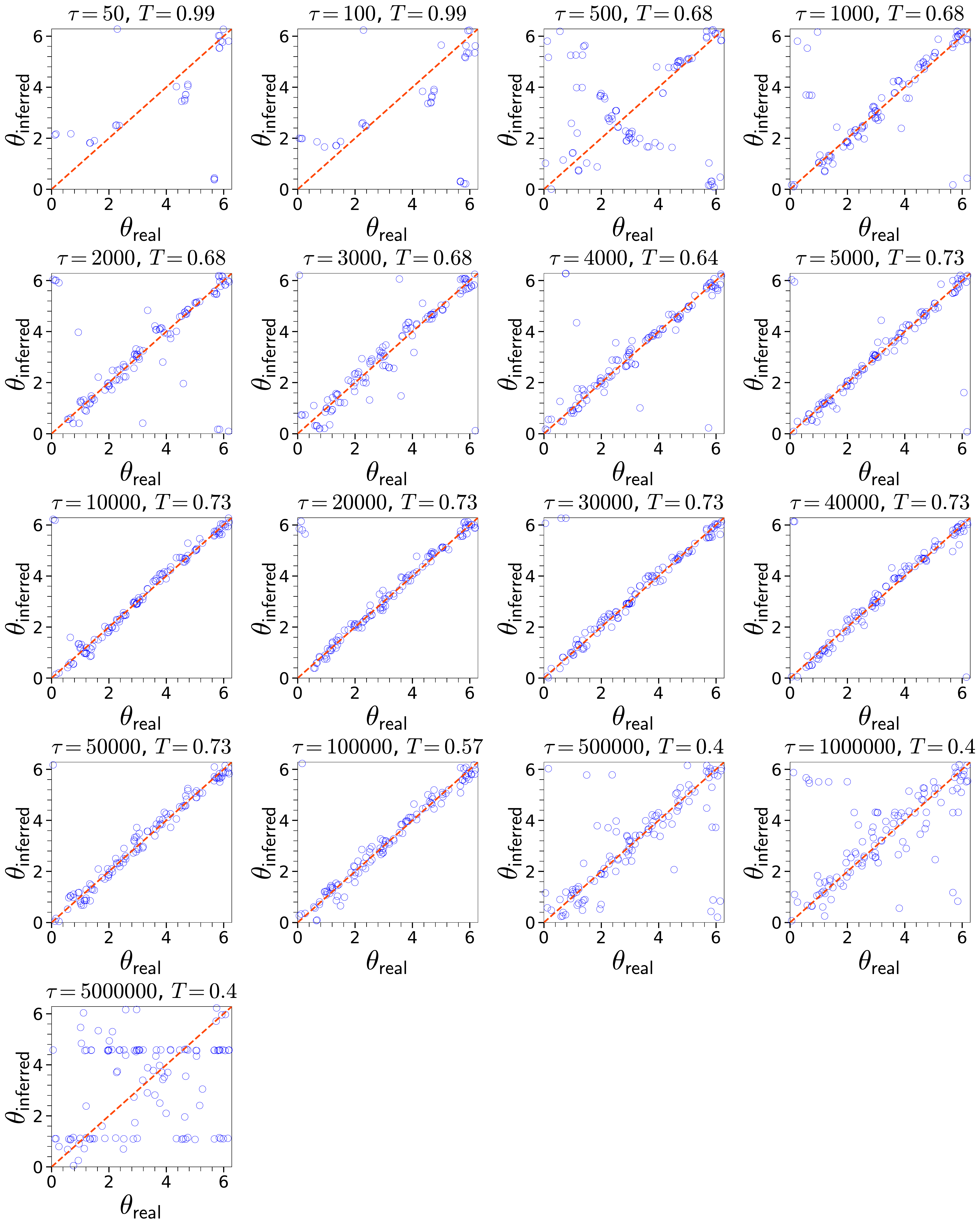}
\caption{\textbf{Inferred~vs.~real $\theta$ for different aggregation intervals $\tau$.} Same as in Fig.~\ref{fig:thetas_diff_intervals_hp} but for the synthetic counterpart of the conference.
\label{fig:thetas_diff_intervals_cf}}
\end{figure}

\begin{figure}[htb!]
\centering
\includegraphics[width=\linewidth]{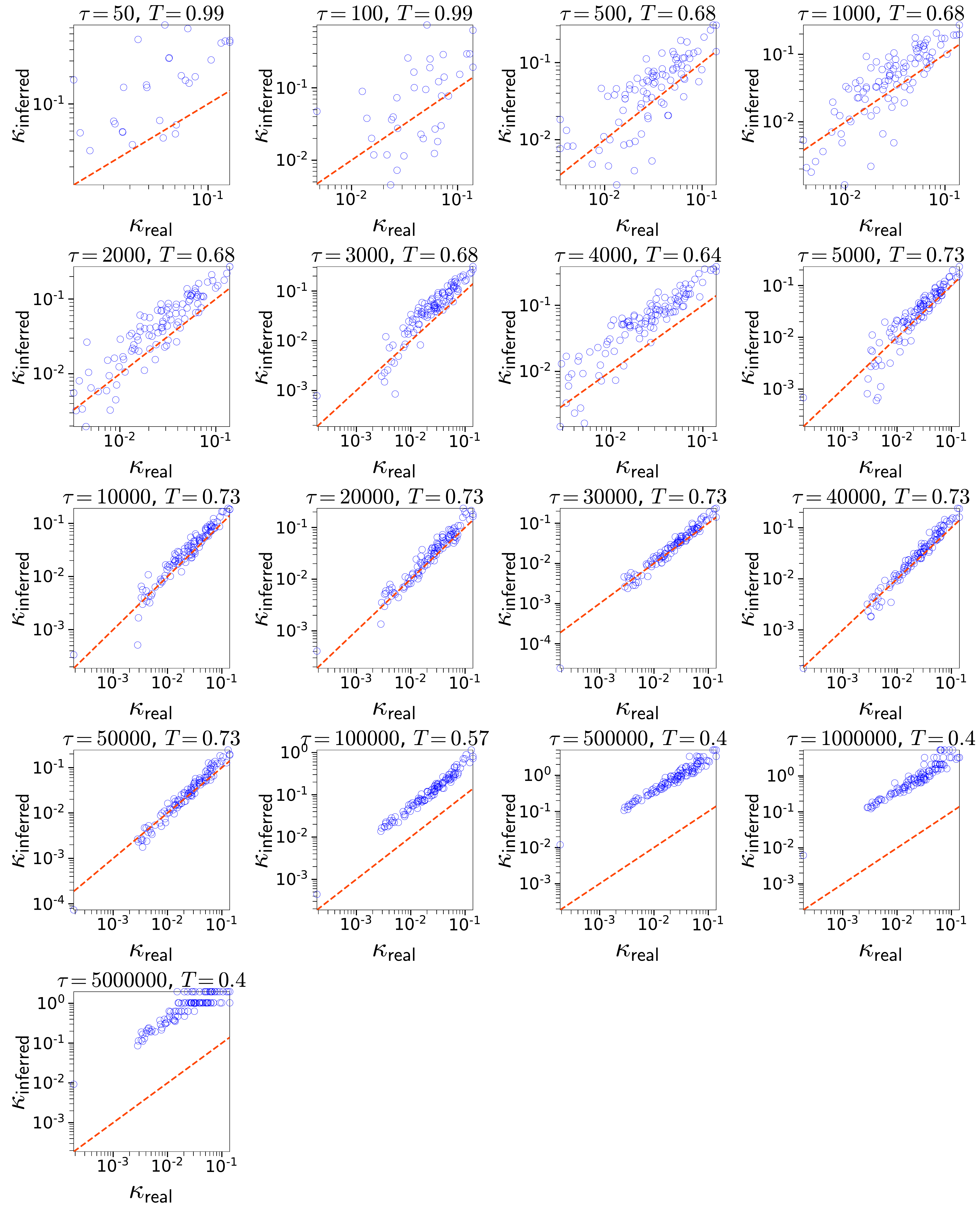}
\caption{\textbf{Inferred~vs.~real $\kappa$ for different aggregation intervals $\tau$.} Same as Fig.~\ref{fig:kappas_diff_intervals_hp} but for the synthetic counterpart of the conference.
\label{fig:kappas_diff_intervals_cf}}
\end{figure}

\begin{figure}[htb!]
\centering
\includegraphics[width=\linewidth]{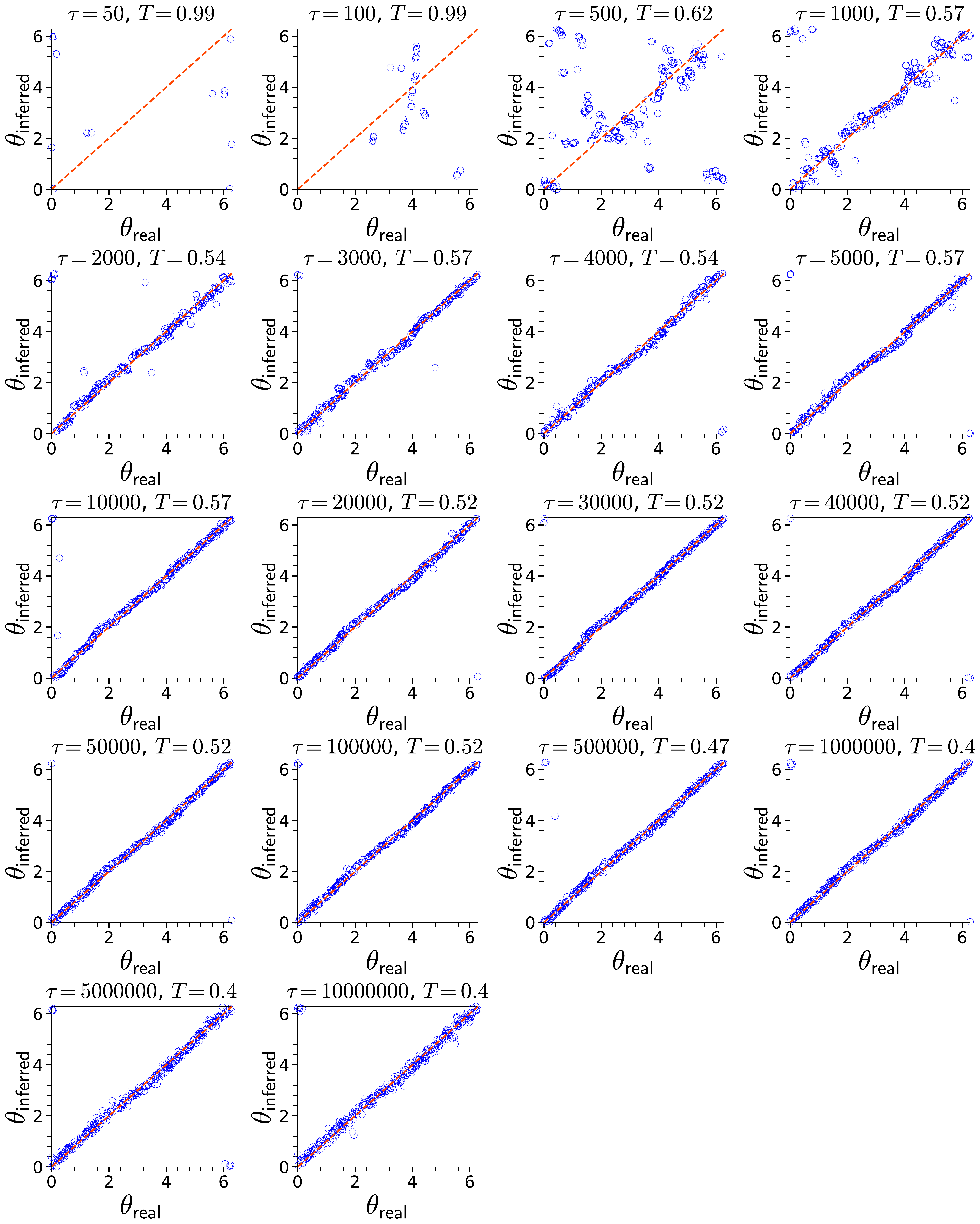}
\caption{\textbf{Inferred~vs.~real $\theta$ for different aggregation intervals $\tau$.} Same as in Fig.~\ref{fig:thetas_diff_intervals_hp} but for the synthetic counterpart of the high school.
\label{fig:thetas_diff_intervals_hs}}
\end{figure}

\begin{figure}[htb!]
\centering
\includegraphics[width=\linewidth]{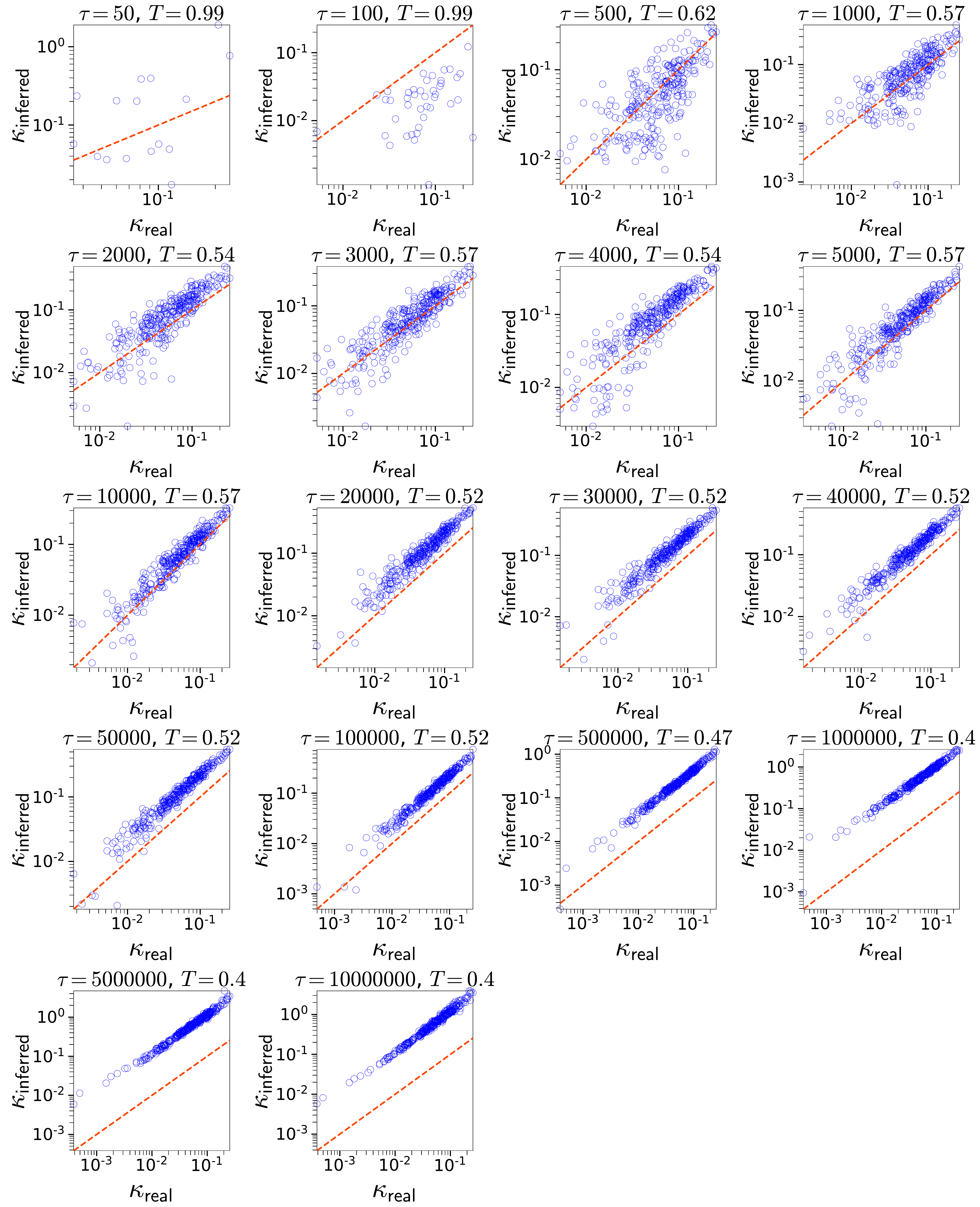}
\caption{\textbf{Inferred~vs.~real $\kappa$ for different aggregation intervals $\tau$.} Same as in Fig.~\ref{fig:kappas_diff_intervals_hp} but for the synthetic counterpart of the high school.
\label{fig:kappas_diff_intervals_hs}}
\end{figure}

\begin{figure}[htb!]
\centering
\includegraphics[width=\linewidth]{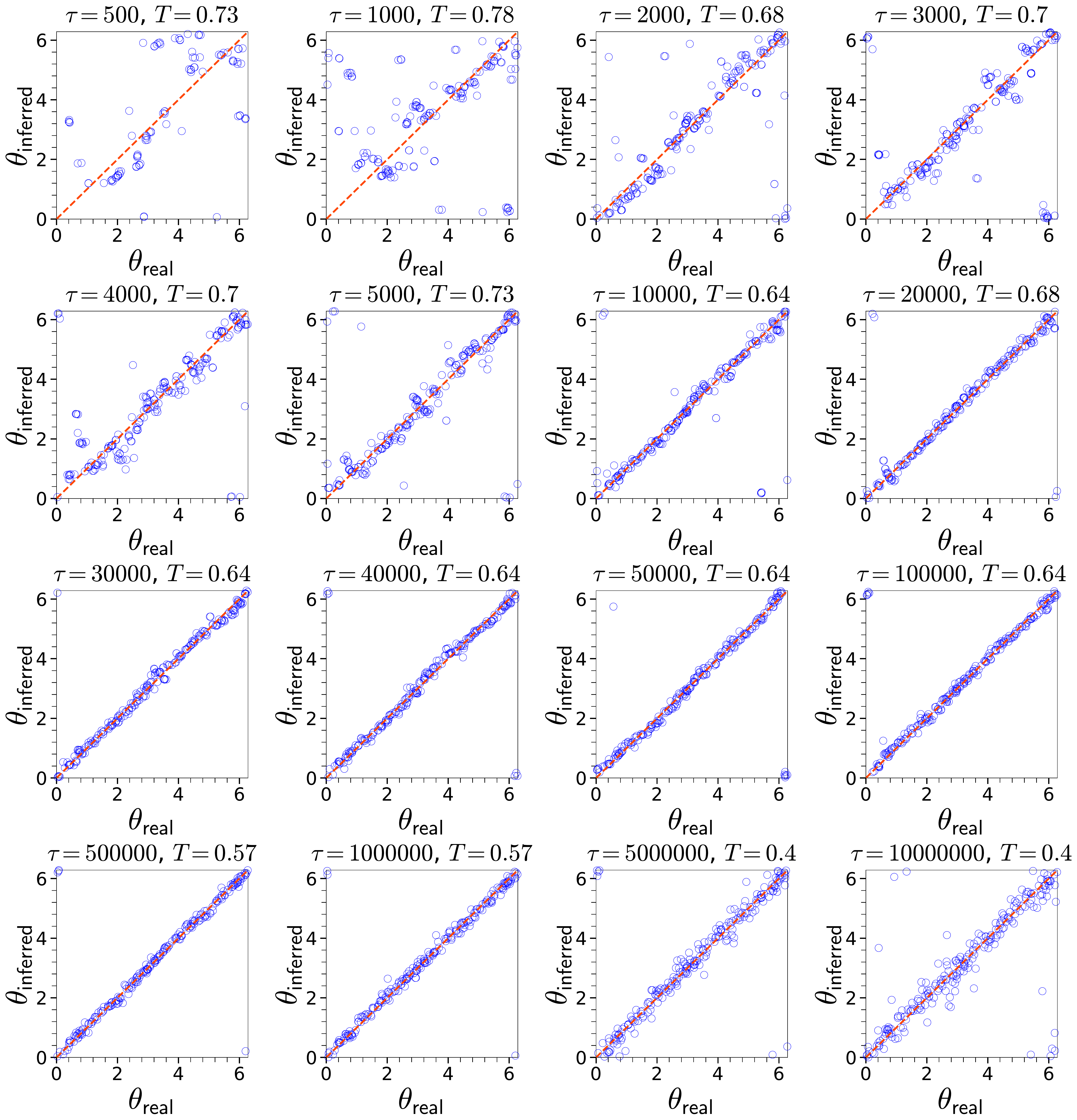}
\caption{\textbf{Inferred~vs.~real $\theta$ for different aggregation intervals $\tau$.} Same as in Fig.~\ref{fig:thetas_diff_intervals_hp} but for the synthetic counterpart of the office building.
\label{fig:thetas_diff_intervals_of}}
\end{figure}

\begin{figure}[htb!]
\centering
\includegraphics[width=\linewidth]{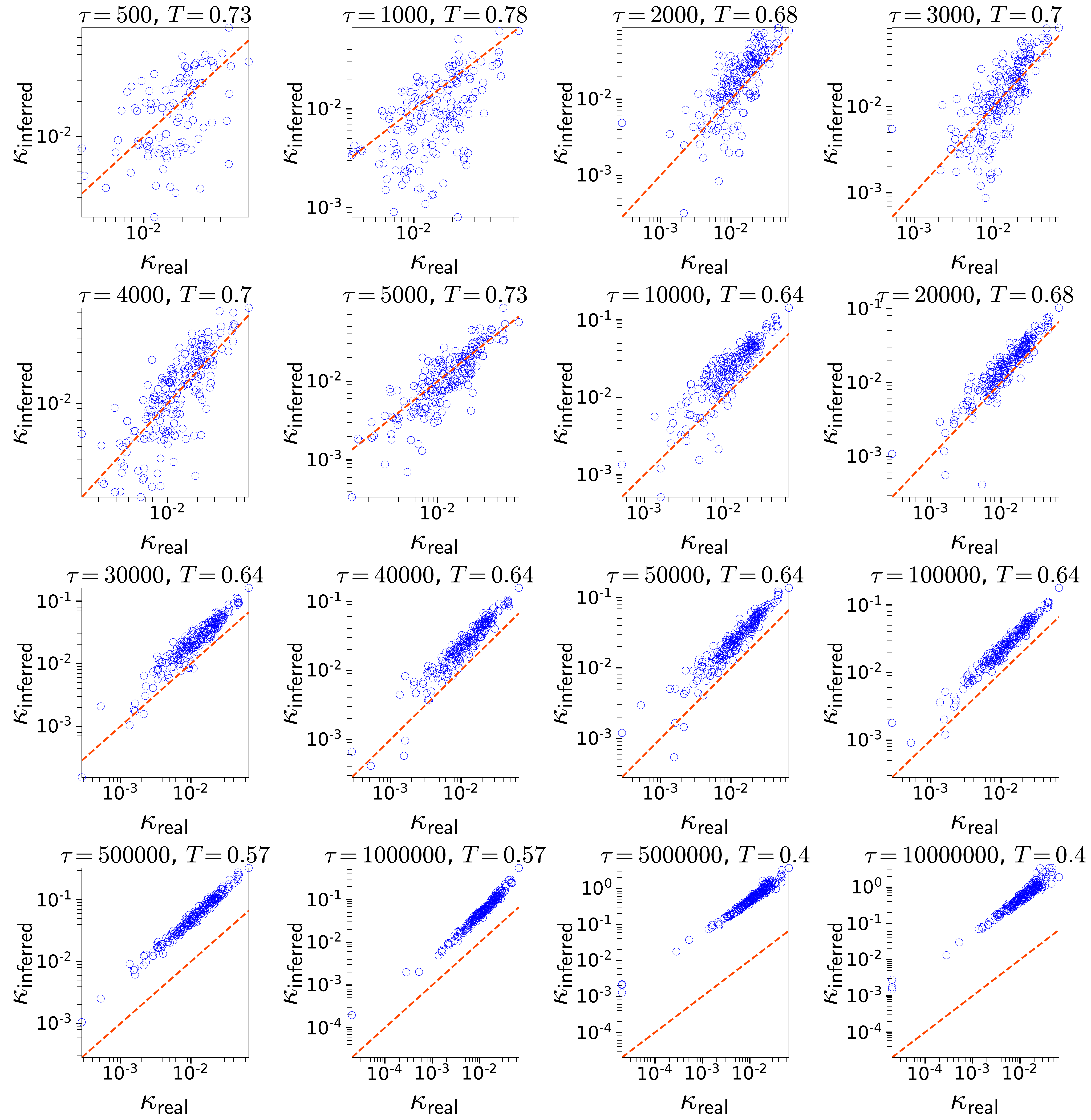}
\caption{\textbf{Inferred~vs.~real $\kappa$ for different aggregation intervals $\tau$.} Same as in Fig.~\ref{fig:kappas_diff_intervals_hp} but for the synthetic counterpart of the office building.
\label{fig:kappas_diff_intervals_of}}
\end{figure}

\begin{figure}[htb!]
\centering
\includegraphics[width=\linewidth]{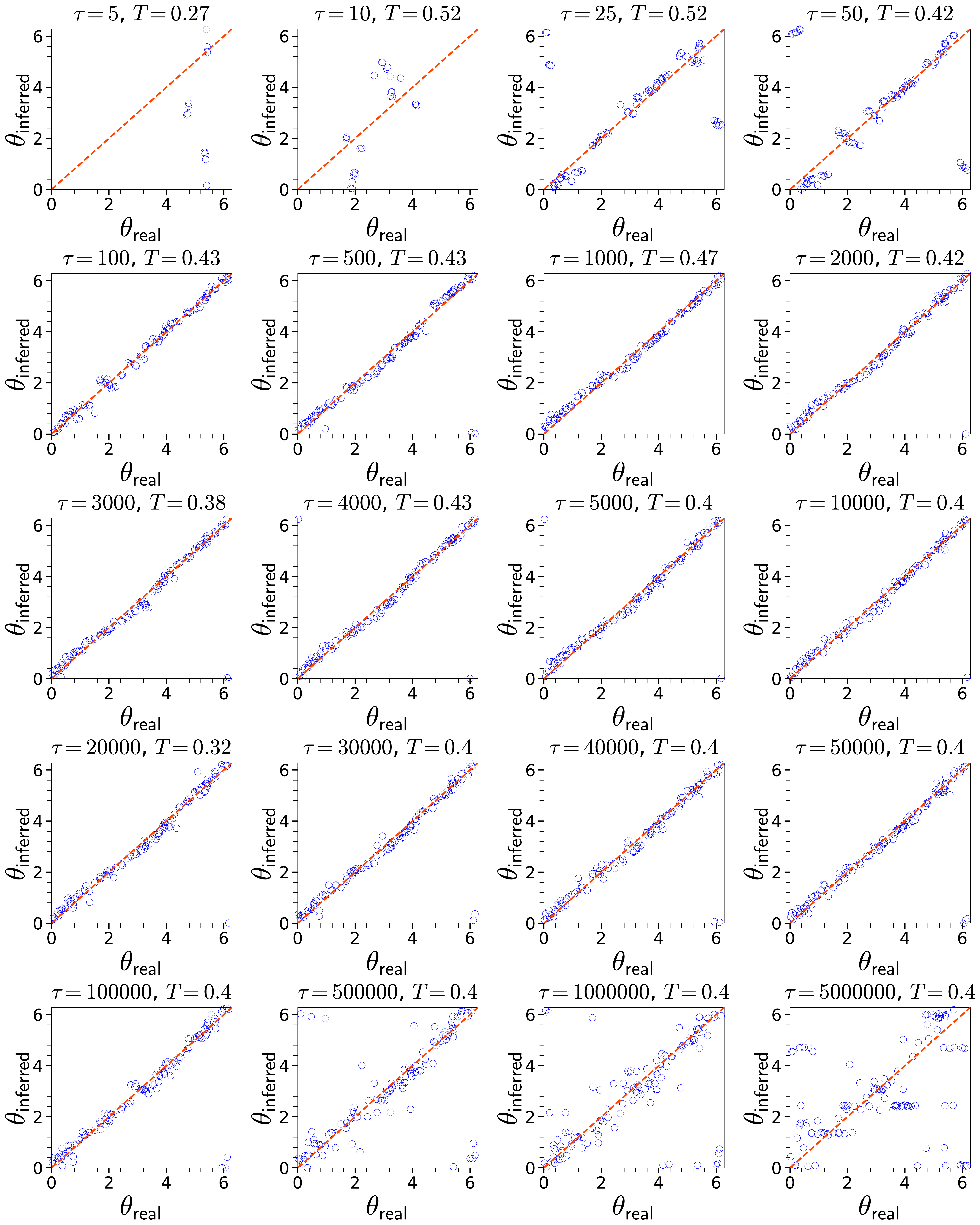}
\caption{\textbf{Inferred~vs.~real $\theta$ for different aggregation intervals $\tau$.} Same as in Fig.~\ref{fig:thetas_diff_intervals_hp} but for the synthetic counterpart of the Friends \& Family.
\label{fig:thetas_diff_intervals_ff}}
\end{figure}

\begin{figure}[htb!]
\centering
\includegraphics[width=\linewidth]{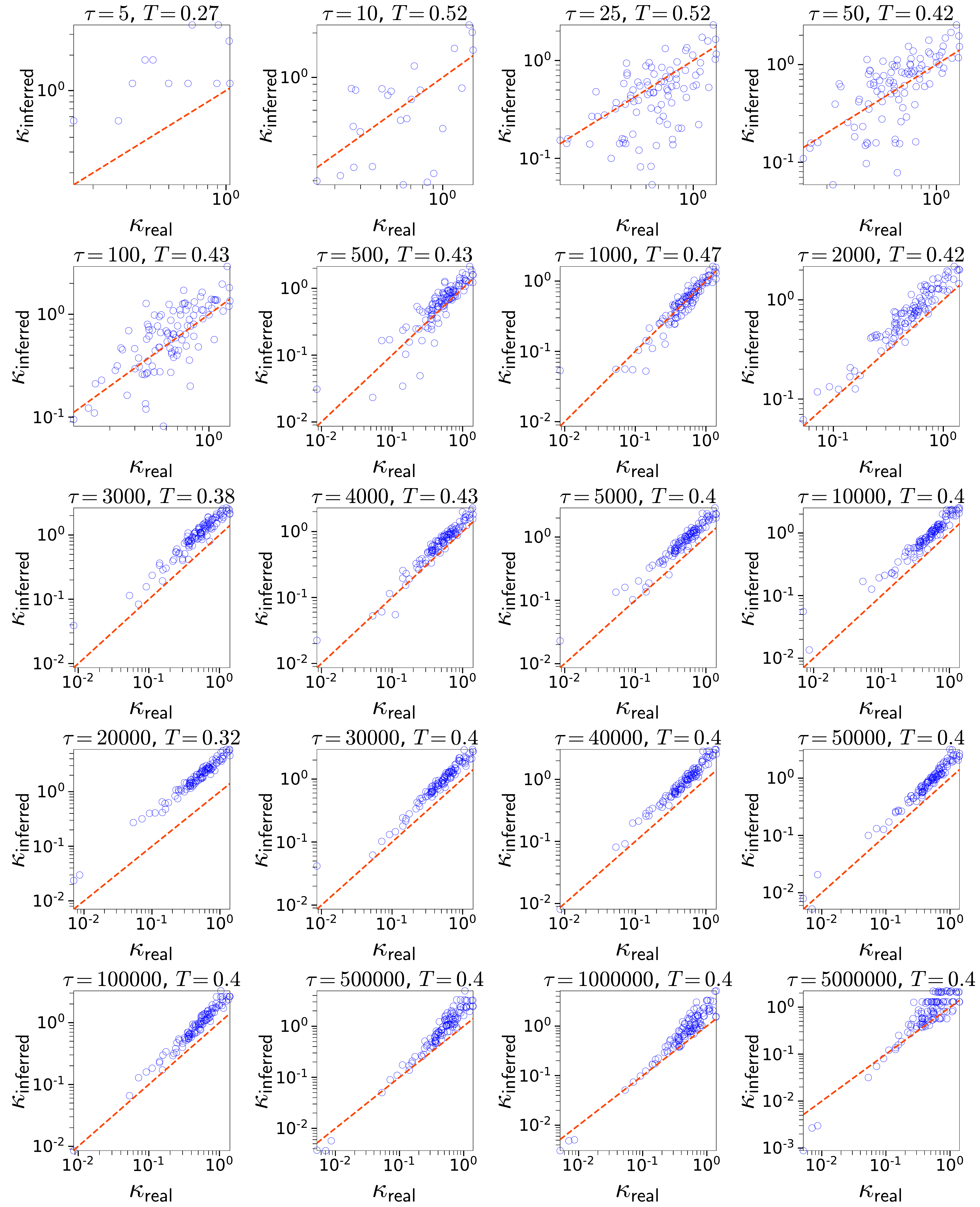}
\caption{\textbf{Inferred~vs.~real $\kappa$ for different aggregation intervals $\tau$.} Same as in Fig.~\ref{fig:kappas_diff_intervals_hp} but for the synthetic counterpart of the Friends \& Family.
\label{fig:kappas_diff_intervals_ff}}
\end{figure}

In Fig.~\ref{fig:aggregation_effect} of the main text we quantify the difference between inferred and real coordinates as a function of the aggregation interval $\tau$ in a synthetic counterpart of the primary school. In this section, Figs.~\ref{fig:interval_hp}-\ref{fig:interval_ff} correspond to the same results but for synthetic counterparts of the hospital, conference, high school, office building and Friends \& Family. Specifically, each of Figs.~\ref{fig:interval_hp}-\ref{fig:interval_ff} shows the metrics $D_{\kappa}(\tau)$, $D_{\theta}(\tau)$ and $d(\tau)$ defined in the caption of Fig.~\ref{fig:aggregation_effect} in the main text. Further, Figs.~\ref{fig:thetas_diff_intervals_hp}-\ref{fig:kappas_diff_intervals_ff} juxtapose the inferred against the real coordinates in each synthetic counterpart, as a function of the aggregation interval $\tau$. 

As mentioned in the main text, before computing $D_{\theta}(\tau)$, we globally shift (rotate) the inferred angles such that the sum of the squared distances (SSD) between real and rotated inferred angles is minimized. To this end, we apply a Procrustean rotation (Ref.~\cite{procrustes_ref}), as follows:

\begin{enumerate}

\item We transform the real and inferred angles $\{\theta_{\textnormal{real}}^{i}\}$ and $\{\theta_{\textnormal{inferred}}^{i}\}$ to Cartesian coordinates $\{x_i, y_i\}=\{\cos{\theta_{\textnormal{real}}^{i}}, \sin{\theta_{\textnormal{real}}^{i}}\}$ and $\{w_i, z_i\}=\{\cos{\theta_{\textnormal{inferred}}^{i}},\sin{\theta_{\textnormal{inferred}}^{i}}\}$ for all nodes $i=1, \ldots, N$.\footnote{Notation``\{~\}" denotes a set. For example, $\{x_i, y_i\}=\{x_1, y_1, x_2, y_2, \ldots, x_N, y_N\}$.}

\item A rotation of the points $\{w_i, z_i\}$ by an angle $\phi$ is given by $\{u_i, v_i\}=\{w_i \cos \phi-z_i \sin \phi, w_i \sin\phi + z_i \cos \phi\}$, where $u_i, v_i$ are the coordinates of the rotated point $w_i, z_i$.  The SSD between $\{u_i, v_i\}$ and $\{x_i, y_i\}$ is $\textnormal{SSD}=\sum_{i=1}^N (u_i-x_i)^2+ (v_i-y_i)^2$. The optimal rotation angle $\phi^{*}$ is computed by taking the derivative of the SSD with respect to $\phi$ and solving for $\phi$ when the derivative is zero,
\begin{equation}
 \phi^{*}=\tan^{-1}\left(\frac{\sum^N_{i=1}(w_i y_i - z_i x_i)}{\sum^N_{i=1}(w_i x_i + z_i y_i)} \right).
\end{equation}

We compute the optimally rotated inferred angles as $\{\theta_{\textnormal{rotated}}^{i}\}=\{\tan^{-1}(v_i^{*}/u_i^{*})\}$, where  $\{u_i^{*}, v_i^{*}\}=\{w_i \cos{\phi^{*}}-z_i \sin{\phi^{*}}, w_i \sin{\phi^{*}} + z_i\cos{\phi^{*}}\}$.\footnote{If  $\theta_{\textnormal{rotated}}^{i} < 0$, then $\theta_{\textnormal{rotated}}^{i} \coloneqq 2 \pi+\theta_{\textnormal{rotated}}^{i}$.}

\item We repeat the above procedure after replacing $\{\theta_{\textnormal{inferred}}^{i}\}$ with $\{2\pi-\theta_{\textnormal{inferred}}^{i}\}$, which is the reflection of the former across the $x$-axis, and compute the optimally rotated inferred angles in this case as well, $\{\tilde{\theta}_{\textnormal{rotated}}^{i}\}$. 

\item We compute $D_\theta(\tau)=\sum_{i=1}^{N}|\theta_{\textnormal{rotated}}^{i}-\theta_{\textnormal{real}}^{i}|/N$ and $\widetilde{D}_\theta(\tau)=\sum_{i=1}^{N}|\tilde{\theta}_{\textnormal{rotated}}^i-\theta_{\textnormal{real}}^{i}|/N$. The optimally rotated inferred angles are  $\{\theta_{\textnormal{rotated}}^{i}\}$ if $D_\theta(\tau) < \widetilde{D}_\theta(\tau)$, and $\{\tilde{\theta}_{\textnormal{rotated}}^i\}$ otherwise.  

\end{enumerate}
We follow a similar procedure for the rotations in Fig.~\ref{fig:stability_maps}.



\clearpage

\end{document}